\documentclass[12pt]{article}

\usepackage{amsmath,amssymb,amsthm,mathrsfs,bbm}
\usepackage{latexsym,amscd,amsbsy,dsfont,amsfonts}
\usepackage{graphicx}
\usepackage[pdftex,colorlinks=true]{hyperref}

\newcommand{\hs}{\hspace{0.15mm}}

\newcommand{\ph}{\phantom{A}}

\newcommand{\bb}{b}

\usepackage{bm}
\usepackage{latexsym}
\usepackage[latin1]{inputenc}
\usepackage{amsfonts}
\usepackage{amssymb}
\usepackage{amsmath}
\usepackage{subfigure}

\numberwithin{equation}{section}

\begin{document}

\title{Schr\"odinger Holography with $z=2$}
\author{Tom\'as Andrade$^{1}$\footnote{tomas.andrade@physics.ox.ac.uk}, Cynthia Keeler$^{2}$\footnote{keeler@nbi.ku.dk}, Alex Peach$^{3}$\footnote{a.m.peach@durham.ac.uk} {} and Simon F. Ross$^{3}$\footnote{s.f.ross@durham.ac.uk} \\  \bigskip \\
 ${}^{1}$Rudolf Peierls Centre for Theoretical Physics, University of Oxford, \\
 1 Keble Road, Oxford OX1 3NP, United Kingdom \\ \bigskip \\
 ${}^{2}$Niels Bohr International Academy,
Niels Bohr Institute \\
University of Copenhagen\\
Blegdamsvej 17,
DK 2100, Copenhagen {\O},
Denmark\\ \bigskip \\
 ${}^{3}$Centre for Particle Theory, Department of Mathematical Sciences \\ Durham University \\ South Road, Durham DH1 3LE}

\maketitle

\begin{abstract}
We investigate holography for asymptotically Schr\"odinger spacetimes, using a frame formalism based on the anisotropic scaling symmetry. We build on our previous work on $z<2$ to propose a dictionary for $z=2$. For $z=2$, the scaling symmetry does not act on the additional null direction, which implies that in our dictionary it does not correspond to one of the field theory directions. This is significantly different from previous analyses based on viewing Schr\"odinger as a deformation of AdS. We study this dictionary in the linearised theory and in an asymptotic expansion. We show that a solution exists in an asymptotic expansion for arbitrary sources for the relevant operators in the stress energy complex.
\end{abstract}

\section{Introduction}

Holography for non-relativistic field theories has been actively studied for several years now. It has the potential to offer us tools to study a broader class of field theories holographically, which may include theories of interest for modelling condensed matter physics \cite{Son:2008ye,Balasubramanian:2008dm,Kachru:2008yh}. It also offers the possibility to deepen our understanding of holographic relations between field theories and gravity. The non-relativistic theories of interest are characterised by the existence of an anisotropic scaling symmetry which treats the time and space directions differently, $
t \to \lambda^z t, \quad x \to \lambda x$,
where $z$ is called the dynamical exponent. There are two cases of interest, Schr\"odinger and Lifshitz, where in the first case the theory has a Galilean boost symmetry, and in the latter there is no such symmetry, so the theory has a preferred rest frame. As a result Schr\"odinger theories have a conserved particle number which is not present in the Lifshitz case. The case $z=2$ is special for Schr\"odinger since in this case the theory has an additional special conformal symmetry. A holographic dual for theories with Schr\"odinger symmetry was proposed first \cite{Son:2008ye,Balasubramanian:2008dm},\footnote{Schr\"odinger space-times also appear in the work of \cite{Duval:2008jg}.} but the Lifshitz case \cite{Kachru:2008yh} has been more fully explored, because of its greater simplicity and close resemblance to the well-understood AdS case. In \cite{Andrade:2014iia}, we used insights from the Lifshitz case to propose a holographic dictionary for Schr\"odinger with $z<2$. In this paper we build on this to discuss the case $z=2$, which has some significant differences.

For a $(d_s+3)$-dimensional Schr\"odinger spacetime, the bulk metric is
\begin{equation} \label{schr}
ds^2 = -\frac{dt^2}{r^{2z}} + \frac{2 dt d\xi + d\vec{x}_{d_s}^2+ dr^2}{r^2},
\end{equation}
where the boundary lies at $r \to 0$ and the number of spatial boundary dimensions is $d_s$. The isometry 
$t \to \lambda^z t$, $x \to \lambda x$, $\xi \to \lambda^{2-z} \xi $, $r \to \lambda r$ realises the anisotropic scaling symmetry, 
and there are isometries $\vec{x} \to \vec{x} + \vec{v} t$, $\xi \to \xi - \vec v \cdot \vec x - \frac{1}{2} v^2 t$,  which realise the 
Galilean boost symmetry. Note that the scaling symmetry  does not act on $\xi$ for $z = 2$. The presence of the additional 
null direction $\xi$ can be understood in field theory terms, as arising from realising the non-relativistic field theory with 
Galilean boosts via the light cone reduction of a Lorentz-invariant theory in one higher dimension \cite{Son:2008ye,Nishida:2007pj}.

By a coordinate transformation $t \to \bb t$, $\xi \to \bb^{-1} \xi$, the metric \eqref{schr} can be rewritten as
\begin{equation}
ds^2 = -\frac{\bb^2 dt^2}{r^{2z}} + \frac{2 dt d\xi + d\vec{x}_{d_s}^2+dr^2}{r^2},
\end{equation}
and for small $\bb$ the geometry outside of some neighbourhood of $r=0$ can be viewed as a deformation of AdS. This motivated the programme of \cite{Guica:2010sw}, which studies Schr\"odinger spacetime holographically as the perturbation of a relativistic theory by an irrelevant vector operator, decomposing the linearised fluctuations of bulk fields in terms of sources and vevs of operators of given scaling dimension with respect to the relativistic scaling symmetry. Continuations and related studies include \cite{Costa:2010cn,Guica:2011ia,vanRees:2012cw}. This programme has had some success, but because the deforming operator is irrelevant, the understanding can only be perturbative in $\bb$.

Our aim is instead to formulate a holographic dictionary based on the anisotropic scaling symmetry, using a frame formalism as in the Lifshitz case \cite{Ross:2009ar,Ross:2011gu}. Such a formulation was attempted for Schr\"odinger in \cite{Hartong:2013cba} for $z=2$, where an appropriate choice of frame fields and boundary conditions was identified, but difficulties were encountered in solving the equations of motion in an asymptotic expansion for general boundary conditions. We will see that these difficulties arose because \cite{Hartong:2013cba} considered sources with arbitrary dependence on $\xi$, which necessarily includes sources for irrelevant operators.  For $z<2$, the anisotropic scaling symmetry and the frame formulation provides a different perspective on the dictionary (the construction of which was carried out in \cite{Andrade:2014iia}) but the discussion can be relatively easily related to the earlier discussion in \cite{Costa:2010cn}.

For $z=2$, however, focusing on the anisotropic scaling symmetry gives a qualitatively different dictionary. In the relativistic theory, the usual scaling symmetry acts non-trivially on all the boundary coordinates, so one thinks of the dual as living in the $(t, \xi, \vec{x})$ space. Bulk fields are dual to local operators $\mathcal O(t,\xi, \vec{x})$. The anisotropic scaling, by contrast, does not act on the $\xi$ direction. Furthermore, as has been known since \cite{Son:2008ye}, the asymptotic behaviour of bulk fields, and hence the scaling dimension of dual operators, depends on $k_\xi$. Hence if we want to focus on the anisotropic scaling symmetry, the natural dual is a field theory living in the $(t, \vec{x})$ space, with local operators $\mathcal O_{k_\xi}(t, \vec{x})$. To relate bulk fields holographically to these operators, we need to expand the bulk fields in Fourier modes in the $\xi$ direction.\footnote{For scalar fields, this is a straightforward Fourier expansion. For tensor fields, we also need to perform a decomposition into components along $\xi$ and in the transverse space. This is complicated by the null nature of $\xi$ in the background \eqref{schr}, as we discuss below.} The situation is analogous to $AdS_2 \times \mathbb R^d$ backgrounds, where the dual is a theory with local operators $\mathcal O_k(t)$ whose dimensions depending on momentum in the spatial directions \cite{Faulkner:2009wj}. The limit as $z \to 2$ from below is analogous to the $z \to \infty$ limit of Lifshitz, which gives the  AdS$_2 \times \mathbb{R}^d$ geometry.

 The Schr\"odinger case is however more complicated than the $AdS_2 \times \mathbb R^d$ case because the metric \eqref{schr} is not a direct product of a scaling and non-scaling part. Moreover, the $\xi$ direction is at least asymptotically null, so we can't decompose the metric in a standard Kaluza-Klein reduction. However, in our holographic context it is more natural for us to think in terms of the one-form frame fields, which we can simply decompose into their component along $d\xi$ and their components along the remaining boundary directions. The zero-modes (under $\partial_\xi$) in the leading terms in the frame fields in the bulk will be interpreted as sources for the stress energy complex in the non-relativistic field theory living in the $(t, \vec{x})$ directions. We will therefore primarily focus on understanding holography for $z=2$ for the sector with $k_\xi =0$, that is, for $\xi$-independent sources. This class includes arbitrary sources for the stress energy complex in the non-relativistic field theory.

We start in the next section by reviewing in more detail the Schr\"odinger metric as a solution of the massive vector theory we will work in for the remainder of this paper (although it should be easy to extend these ideas to alternative realizations of Schr\"odinger such as topologically massive gravity), and recalling the frame boundary conditions introduced in \cite{Andrade:2014iia}. In section \ref{kk}, we discuss the analogue of Kaluza-Klein reduction in our frame formalism. In section \ref{stress}, we review the structure of the stress energy complex for non-relativistic theories. In section \ref{lin} we set up the linearised analysis around the Schr\"odinger solution for $z=2$ in general dimensions. Section \ref{spat} discusses the case of two boundary spatial dimensions, $d_s=2$, identifying the linearised modes with sources and vevs for the stress energy complex. Section \ref{dszero} discusses the special case $d_s=0$, including its degenerate Ward identities, and compares it to previous work. In section \ref{asymp}, we discuss the asymptotic expansion for $z=2$, and show that a solution can be obtained in an expansion in powers of $r$,\footnote{In our analysis, this is traded for an expansion in eigenvalues of a suitable dilatation operator, but the existence of a dilatation expansion implies the existence of an expansion in powers of $r$, since each term in the dilatation expansion has an expansion in positive powers of $r$.}  and that all divergences in the action can be eliminated by adding boundary counterterms which are local functions of the boundary data. We summarise and discuss future directions in section \ref{disc}. 

\section{Asymptotically locally Schr\"odinger boundary conditions}
\label{bc}

We consider the metric \eqref{schr} as a solution of the theory with a massive vector introduced in \cite{Son:2008ye}. The action is
\begin{equation}
S =-\frac{1}{16 \pi G} \int d^{d_s + 3} x \sqrt{-g} \left(R -2 \Lambda - \frac{1}{4} F_{\mu\nu} F^{\mu\nu} - \frac{1}{2} m^2 A_\mu A^\mu \right) - \frac{1}{8 \pi G} \int d^{d_s +2} x \sqrt{-h} K,
\end{equation}
with 
\begin{equation}
	m^2 = z(z+d_s), \quad \Lambda = - \frac{(d_s+2)(d_s+1)}{2}.
\end{equation}
Note $d_s$ labels the number of boundary spatial directions. The equations of motion that follow are
\begin{equation}\label{eins eqs}
	R_{\mu \nu} - \frac{1}{2} R g_{\mu \nu} + \Lambda g_{\mu \nu}
	= \frac{1}{2} \left(  F^\rho \hs _\mu F_{\rho \nu} - \frac{1}{4} F^2 g_{\mu \nu}   \right)
	+ \frac{m^2}{2}\left(  A_\mu A_\nu - \frac{1}{2} A^2 g_{\mu \nu}  \right),
\end{equation}
\begin{equation}\label{maxw eq}
	\nabla_\mu F^{\mu \nu} = m^2 A^\nu.
\end{equation}
The metric \eqref{schr} is a solution of \eqref{eins eqs}, \eqref{maxw eq} supported by the matter field
\begin{equation}\label{A sch}
	A = \alpha  r^{-z} dt , \quad \alpha  = \sqrt{\frac{2(z-1)}{z}}.
\end{equation}
The massive vector field $A_\mu$ physically singles out the $t$ direction as special. We will henceforth set $z=2$, which implies $\alpha=1$.

We want to define a class of asymptotically locally Schr\"odinger spacetimes which asymptotically approach \eqref{schr} locally as $r \to 0$. Inspired by the analysis in the Lifshitz case \cite{Ross:2009ar,Ross:2011gu} and previous discussion of Schr\"odinger in  \cite{Hartong:2013cba}, in \cite{Andrade:2014iia} we proposed that the asymptotics are defined in terms of a set of frame fields $e^A$, $A= +,-,I,r$ such that the metric is
\begin{equation} \label{a metric}
ds^2 = g_{AB} e^A e^B =  - e^+ e^+ + 2 e^+ e^- + e^I e^I + e^r e^r.
\end{equation}
We will always adopt the radial gauge choice $e^r = r^{-1} dr$. Consequently, in subsequent expressions, $A$ runs over only $+,-,I$; $I$ runs over the $d_s$ spatial frame indices. In the background \eqref{schr} at $z=2$, $e^+ = r^{-2} dt$, $e^- =  d\xi$, $e^I = r^{-1} dx^i$, so each of the frame fields has a well-defined scaling with $r$ at small $r$ (i.e near the boundary).\footnote{Note that for this flat background, the frame index $I$ and the coordinate index $i$ are equivalent.} Note that in the present $z=2$ case, $e^-$ does not scale with $r$. This is the bulk expression of the invariance of $\xi$ under the anisotropic scaling. We further restrict the frame fields by assuming that the vector field can be written as
\begin{equation} \label{a vector}
A = e^+ + \psi e^-  + s_r e^r,
\end{equation}
where $\psi$ is the single scalar degree of freedom in the boundary conditions for the matter field; $s_r$ labels the radial component of the field which is left arbitrary here. We will find that the operator dual to $\psi$ is irrelevant, so we always set the source part to zero. 

There is a residual gauge symmetry in the choice of frames which consists of
\begin{equation} \label{res}
e^+ \to e^+ - \psi \beta^I e^I, \quad e^- \to e^- + \beta^I e^I, \quad e^I \to e^I - \beta^I e^+ - \psi \beta^I (e^+ - e^-),
\end{equation}
together with the rotations of the spatial frame fields $e^I$.

In \cite{Andrade:2014iia}, a spacetime was defined to be asymptotically locally Schr\"odinger if the metric and massive vector can be written as in (\ref{a metric}, \ref{a vector}) with
\begin{equation} \label{ALSznot2}
e^+ = r^{-z} \hat e^+, \quad e^- = r^{z-2} \hat e^-, \quad e^I = r^{-1}  \hat e^I,
\end{equation}
and the scalar $\psi = r^{\Delta_-} \hat \psi$ for some exponent $\Delta_-$, where the fields $\hat e^A, \hat \psi$ are arbitrary functions of $t, \xi, \vec{x}, r$ with finite limits as $r \to 0$. Note we do not directly impose a boundary condition on $s_r$, since it does not represent an independent degree of freedom; it is determined algebraically by the other components. 

This definition requires modification for the case of $z=2$. As we argued in the introduction, we think of the dual field theory as living in just the $t, \vec{x}$ directions. As a result, it is just the Fourier zero modes of the frame fields that we expect to provide geometrical boundary data, that is the sources for the dual stress tensor complex living in the $t, \vec x$ directions. We therefore say that a spacetime is {\it  asymptotically locally Schr\"odinger} for $z=2$ if the Fourier zero modes of the frame fields satisfy
\begin{equation} \label{ALS}
e^+_{k_\xi =0} = r^{-2} \hat e^+, \quad e^-_{k_\xi =0} = \hat e^-, \quad e^I_{k_\xi=0}  = r^{-1}  \hat e^I. 
\end{equation}
The non-zero modes of the frame fields will have fall-offs that depend on the momentum $k_\xi$ in the $\xi$ direction (we will see this explicitly in the linearised analysis in section \ref{lin}). These will be dual to some additional tensor operators in the field theory. We do not make any assumption about the boundary conditions for these fields in defining our asymptotically locally Schr\"odinger boundary conditions, but clearly some of them will be irrelevant operators, and to satisfy our boundary condition \eqref{ALS} we will need to set the sources for the irrelevant operators to zero. 

\subsection{Kaluza-Klein decomposition}
\label{kk}

Since we want to relate the bulk theory to a boundary theory living just in the $t, \vec x$ directions, it is useful to set up a decomposition of the bulk fields in the analogue of Kaluza-Klein reduction on the $\xi$ direction. It is natural to decompose the frame fields as
\begin{equation}
\hat e^A = \hat e^A_a dx^a + \hat e^A_\xi d\xi,
\end{equation}
where $a$ runs over $t, x^i$. This decomposition is the analogue in our frame language of the Kaluza-Klein decomposition of the metric. Each of these components should then be expanded in Fourier components with respect to $\xi$.

In the bulk, there are diffeomorphisms which preserve our choice of radial gauge, generated by the vector field
\begin{equation}
\chi = \chi^\alpha \partial_\alpha + \sigma r \partial_r - \frac{1}{2} r^2 \partial_i \sigma \partial_i -\frac{1}{2} r^2 \partial_\xi \sigma \partial_t - \frac{1}{2} ( r^2 \partial_t \sigma + \ln r \partial_\xi \sigma) \partial_\xi,
\end{equation}
where $\chi^\alpha$, $\sigma$ are functions of the boundary coordinates $t, \vec x, \xi$.  We use $\alpha$ to denote all nonradial spacetime coordinates; thus $\alpha$ runs over $t,x^i,\xi$. These generate the diffeomorphisms $\chi^\alpha$ of the boundary coordinates, which act on the boundary frame fields by the Lie derivative $\delta \hat e^A = \mathcal L_\chi \hat e^A$, and an anisotropic Weyl transformation $\sigma$ on the boundary, which acts as $\delta \hat e^+ = 2 \sigma \hat e^+, \delta \hat e^- = 0, \delta \hat e^I = \sigma \hat e^I$. In the context of our Kaluza-Klein decomposition, it is natural to decompose $\chi^\alpha$ and $\sigma$ in Fourier modes in $\xi$. The zero modes in the Fourier decomposition of $\chi^a$, $\sigma$ correspond to diffeomorphisms and a Weyl transformation of the field theory background, while that of $\chi^\xi$ is naturally interpreted as a gauge transformation of the vector $\hat e^-_a$, ensuring that the dual operator is indeed a conserved current.

For the non-zero modes, in a Kaluza-Klein reduction the usual approach is to gauge-fix them. That is, since we are singling out a direction to reduce along, it is natural to use a formalism which is covariant in the lower dimensional space, but where we fix symmetries which depend on the additional direction. In the usual Kaluza-Klein reduction, where we split the metric into a lower-dimensional metric, vector field and scalar, the usual gauge fixing is to set the non-zero modes of the vector and scalar to zero, so that the physical content is a massive tensor field. Analogously, in our frame based description, we will gauge fix the diffeomorphisms $\chi^\alpha$ by setting the non-zero modes of $\hat e^A_\xi$ to zero. That is, we use the diffeomorphism symmetry to make the components along the extra dimension constant in $\xi$. We also have non-zero modes in the Weyl scaling $\sigma$, but we will not gauge fix this as the scaling symmetry may develop an anomaly. 

The zero modes of $\hat e^A_a$ for $A= +, I$ then define the boundary geometry, which provides a background for the dual field theory living in the $t, \vec x$ directions, and will correspond to the sources for the stress complex in the field theory, which will be reviewed in the next subsection. The zero mode of $\hat e^-_a$ is a one-form vector field which provides the source for the conserved current associated to particle number. The zero modes of $\hat e^A_\xi$, which were interpreted as geometrical in the relativistic context, and in our previous discussion for $z<2$, are in this decomposition instead just sources for additional scalar operators, as is $\hat \psi$.\footnote{The $\hat e^I_\xi$ are sources for operators which are scalars under coordinate transformations, but vectors under frame rotations.} After the gauge-fixing above, the non-vanishing components for the $\xi$-dependent modes are the $\hat e^A_a$, which are sources for additional massive vector operators in the dual field theory, with $k_\xi$-dependent dimensions. Our analysis will focus mainly on the zero-modes, as in the usual Kaluza-Klein decomposition. 

\subsection{Stress energy complex}
\label{stress}

We now review the structure of the stress energy complex in a non-relativistic theory.\footnote{In the $z <2$ discussion in \cite{Andrade:2014iia}, we were interested in the stress complex in a higher-dimensional spacetime with $t, \xi, \vec x$ coordinates, but here we are interested in a dual living just in $t, \vec x$, so the relevant stress tensor complex operators are those appearing in \eqref{nre}, \eqref{nrp}, and \eqref{nrrho} below.} Any non-relativistic theory, Lifshitz or Schr\"odinger, will have an energy density $\mathcal E$ and an energy flux $\mathcal E^i$, satisfying the conservation equation (in a flat boundary space) 
\begin{equation} \label{nre}
\partial_t \mathcal E + \partial_i \mathcal E^i =0,
\end{equation}
along with a momentum density $\mathcal P_i$ and a spatial stress tensor $\Pi_{ij}$ satisfying the conservation equation
\begin{equation} \label{nrp}
\partial_t \mathcal P_i + \partial_j \Pi_i^j = 0.
\end{equation}
The Schr\"odinger theory additionally has a conserved particle number, so there is a particle number density $\rho$ and a particle number flux $\rho^i$ satisfying
\begin{equation} \label{nrrho}
\partial_t \rho + \partial_i  \rho^i = 0.
\end{equation}
The scale invariance for $z=2$ implies $2 \mathcal E + \Pi_i^i =0$. $\mathcal E$ has dimension $2+d_s$, which implies $\mathcal E^i$ has dimension $3 +d_s$, and $\mathcal P_i$ has dimension $1+d_s$, which implies $\Pi_{ij}$ has dimension $2+d_s$. The particle number has dimension zero, so its density $\rho$ has dimension $d_s$, so $\rho^i$ has dimension $1+d_s$. In fact, in a non-relativistic theory $\rho^i = \mathcal P_i = \rho v^i$, where $v^i$ is the local velocity of the particles, so $\rho^i$ and $\mathcal P_i$ are not independent operators.

The sources for these operators are then the zero-mode components of the frame fields, $\hat e^A_a$: the components of $\hat e^+$ provide the sources for $\mathcal E, \mathcal E^i$; the components of $\hat e^-$ provide the sources for $\rho, \rho^i$; and the components of $\hat e^I$  provide the sources for $\mathcal P_i$, $\Pi_i^j$. We can think of these as components of the non-symmetric tensor
\begin{equation} 
T^{a}_{\ \ B} = \frac{1}{\sqrt{-h}}\frac{\delta}{\delta e_{a}^{B}}S.
\end{equation}
The residual gauge symmetry \eqref{res} corresponds to the fact that there are not independent physical sources for $\mathcal P_i$ and $\rho^i$, while the symmetry under rotations of the $\hat e^I$ corresponds to the symmetricity of $\Pi_{ij}$.

There are scalar operators whose sources are $\hat e^A_\xi$. Because of the relation to the higher dimensional stress complex, it is natural to refer to these as $\mathcal E^\xi$, $\rho^\xi$, and $\mathcal P_i^\xi$ for $A = +, -, I$ respectively, but we stress again that these are not components of the stress complex in the dual field theory; they are just some scalar operators (with respect to coordinate transformations; $\mathcal P^\xi_i$ is a vector with respect to frame rotations).  They have dimensions $4 + d_s$, $2+d_s$, and $3+d_s$ respectively.

As in the Lifshitz case, there are irrelevant operators in the stress energy complex, and we would expect to need to set their sources to zero. The marginal dimension is $2 + d_s$, so $\mathcal E^i$, $\mathcal E^\xi$ and $\mathcal P_i^\xi$ are irrelevant operators, and as we will see in the linearised analysis in the next subsection, so is the operator $O_\psi$ dual to the scalar source $\psi$. For $z=2$ the asymptotic expansion only exists if we set the sources for all these operators to zero. For the scalar operators $O_\psi$, $\mathcal E^\xi$ and  $\mathcal P_i^\xi$ we need to set the sources to zero by hand. This implies that of the frame fields, only $e^-$ is allowed to have a leading component along $d\xi$. The only irrelevant operator in the stress tensor complex is $\mathcal E^i$. We can set its source to zero by adopting the irrotational condition
\begin{equation}\label{irrotcond}
\hat e^+ \wedge d \hat e^+= 0.
\end{equation}
As in Lifshitz, this can be viewed as a condition that the boundary geometry defined by the $\hat e^A$ admits a foliation by surfaces of absolute time, as is appropriate for a non-relativistic theory.

\section{Linearised analysis: generalities}
\label{lin}

We now turn to a linearised analysis of the equations of motion (\ref{eins eqs},\ref{maxw eq}) for $z=2$. This will enable us to confirm several of the features we have asserted in our discussion so far: we will see how the scaling behaviour of bulk fields depends on the momentum $k_\xi$, and we will see that the set of solutions for $k_\xi = 0$ has the expected structure to correspond to the stress tensor complex and its sources. In this section, we set up the general formalism. In the following two sections we will discuss $d_s=2$ and $d_s=0$ in detail.

The linearised version of our frame fields is
\begin{eqnarray}
\hat e^+ &=& (1 + \delta \hat e^+_t) dt +  \delta \hat e^+_\xi d\xi +  \delta \hat e^+_i dx^i, \\
\hat e^- &=& (1 + \delta \hat e^-_\xi) d\xi + \delta \hat e^-_t dt +  \delta \hat e^-_i dx^i, \\
\hat e^I &=&  (\delta^I_j + \delta \hat e^I_j) dx^j + \delta \hat e^I_t dt +  \delta \hat e^I_\xi d\xi.
\end{eqnarray}
The linearised fields are then $\delta \hat e^A_\alpha$ and the $\psi$, $s_r$ in \eqref{a vector}. The zero modes in $\delta \hat e^A_a$ are assumed to represent sources for the corresponding components of the stress tensor complex. The zero modes in $\delta \hat e^A_\xi$ are sources for scalar operators.

The linearised version of the residual gauge symmetry \eqref{res} is $ \delta \hat e^-_i \to \delta \hat e^-_i + \hat \beta^i $, $\delta \hat e^I_t \to \delta \hat e^I_t - \hat \beta^i $ (where $\beta^I = r \hat \beta^i$). This implies that the sources for $\mathcal P_i$ and $\rho^i$ are not independent, as expected. The rotation symmetry of the $e^I$ also implies that only the symmetric part of $\delta e^I_j$ provide independent sources. The equations of motion are easier to discuss in the metric language, so we will resolve this gauge symmetry by passing back from the frame fields to the metric and vector for this linearised analysis.

In the metric language, the linearised perturbations are $h_{\mu\nu}$, $a_\mu$.
The linearised equations in the metric language are as in \cite{Ross:2009ar}\footnote{Note that $h_{\mu\nu}$ denotes
  the perturbation of the metric, and indices are raised and lowered
  with the background metric, so $h^{\mu\nu}$ is the perturbation of
  the metric with the indices raised, not the perturbation of the
  inverse metric.}
\begin{equation}
\nabla_\mu f^{\mu\nu} - \nabla_\mu (h^{\mu\lambda} F_\lambda^{\ \nu})
- \nabla_\mu h^{\beta \nu} F^\mu_{\ \beta}
+ \frac{1}{2} \nabla_\lambda h F^{\lambda \nu} = m^2 a^\nu
\end{equation}
and
\begin{eqnarray}
R_{\mu\nu}^{(1)} &=& \frac{2}{d-2} \Lambda h_{\mu\nu} + \frac{1}{2} f_{\mu\lambda} F_\nu^{\ \lambda} +
\frac{1}{2} f_{\nu\lambda} F_\mu^{\ \lambda} - \frac{1}{2}
F_{\mu\lambda} F_{\nu\sigma} h^{\lambda \sigma} -\frac{1}{2(d-2)}
f_{\lambda \rho} F^{\lambda \rho} g_{\mu\nu}   \nonumber \\ &&+ \frac{1}{2(d-2)} F_{\lambda
  \rho} F_\sigma^{\ \rho} h^{\lambda \sigma} g_{\mu\nu}- \frac{1}{4(d-2)}
F_{\lambda \rho} F^{\lambda \rho} h_{\mu \nu} + \frac{1}{2} m^2
a_\mu A_\nu + \frac{1}{2} m^2 a_\nu A_\mu,
\end{eqnarray}
where $d=d_s+3$ is the dimension of the spacetime, $f_{\mu\nu} = \partial_\mu a_\nu - \partial_\nu a_\mu$ and
\begin{equation}
R_{\mu\nu}^{(1)} = \frac{1}{2} g^{\lambda \sigma} [ \nabla_\lambda
  \nabla_\mu h_{\nu \sigma} + \nabla_\lambda \nabla_\nu h_{\mu\sigma} -
  \nabla_\mu \nabla_\nu h_{\lambda \sigma} - \nabla_\lambda
  \nabla_\sigma h_{\mu\nu} ].
\end{equation}

It is convenient to write
\begin{equation}
	h_{tt} = r^{-4} H_{tt},  \qquad h_{t \xi} = r^{-2} H_{t \xi},  \qquad h_{\xi \xi} =  H_{\xi \xi},
\end{equation}
\begin{equation}
	h_{ti} =  r^{-3} H_{ti} \qquad h_{\xi i} = r^{-1} H_{\xi i}, \qquad h_{ij} = r^{-2} H_{ij},
\end{equation}
\begin{equation}
	a_r = r^{-1} s_r  \qquad a_t = r^{-2} s_t  \qquad a_\xi = s_\xi  \qquad  a_{i} = r^{-1} s_{i}.
\end{equation}
Then a given linearised mode will contribute at the same order in $r$ in all the different fields, and the power of $r$ will correspond to the scaling dimension of the mode. The $s_r$ here is the same as in \eqref{a vector}, and the other fields are related to the linearised frame fields by
\begin{equation} \label{lin map 1}
H_{tt} = -2 \delta \hat e^+_t + 2 r^{2} \delta \hat e^-_t, \quad H_{t\xi} = - r^{-2} \delta \hat e^+_\xi  + \delta \hat e^-_\xi + \delta \hat e^+_t, \quad H_{\xi\xi} = 2 r^{-2} \delta \hat e^+_\xi,
\end{equation}
\begin{equation}\label{lin map 2}
H_{ti} = - r^{-1} \delta \hat e^+_i + r \delta \hat e^-_i + r \delta \hat e^I_t, \quad H_{\xi i} = r^{-1} \delta \hat e^+_i + r^{-1} \delta \hat e^I_\xi, \quad H_{ij} =  \delta \hat e^I_j + \delta \hat e^J_i,
\end{equation}
\begin{equation} \label{lin map 3}
s_t = \delta \hat e^+_t, \quad s_\xi = r^{-2} \delta \hat e^+_\xi + \psi, \quad s_i = r^{-1} \delta \hat e^+_i.
\end{equation}
In \eqref{lin map 2} and where appropriate subsequently, the reader should understand $\delta\hat{e}_t^I$ in $H_{ti}$ to stand for $\delta\hat{e}_t^I \delta_{Ii}$, where $\delta_{Ii}$ is the Kronecker delta, and similarly for other components with an $i$ index. 

\subsection{Flux and inner product}

We will identify the modes corresponding to components of the stress tensor by adopting the approach of  \cite{Papadimitriou:2010as},  computing the symplectic flux at the boundary $r=0$, and identifying the modes canonically conjugate to the sources with the vevs. The appropriate symplectic current for the Einstein-massive vector theory we are considering was worked out in  \cite{Andrade:2013wsa}. It involves combining the usual gravitational symplectic current $j^\mu_g$ with an additional component from the massive vector field, $j^\mu_a$:
\begin{equation}
	j^\mu = j^\mu_g + j^\mu_a.
\end{equation}
These are respectively given by
\begin{equation}
j^\mu_g(h_{\bf 1}, h_{\bf 2}) = P^{\mu\nu\alpha\beta\gamma\delta} (h_{ \bf 2 \alpha \beta}^* \nabla_\nu h_{\bf 1 \gamma \delta} 
- h_{\bf 1 \alpha \beta} \nabla_\nu h_{\bf 2 \gamma \delta}^*),
\end{equation}
\begin{equation}
j^\mu_a( \{h_{\bf 1},  a_{\bf 1} \}, \{ h_{\bf 2}, a_{\bf 2} \} ) = a_{\bf 2 \nu}^* (f_{\bf 1}^{\mu \nu} - h_{\bf 1}^{\mu\lambda} F_\lambda^{\ \nu} 
- h_{\bf 1}^{\beta \nu} F^\mu_{\ \beta} + \frac{1}{2} h_{\bf 1} F^{\mu \nu}) - (\bf 1 \leftrightarrow \bf 2),
\end{equation}
where
\begin{equation}
P^{\mu\nu\alpha\beta\gamma\delta} = \frac{1}{2} (g^{\mu\nu} g^{\gamma (\alpha} g^{\beta) \delta} + g^{\mu (\gamma} g^{\delta) \nu} g^{\alpha \beta} + g^{\mu (\alpha} g^{\beta) \nu} g^{\gamma \delta} - g^{\mu\nu} g^{\alpha \beta} g^{\gamma \delta} - g^{\mu (\gamma} g^{\delta) (\alpha} g^{\beta) \nu} - g^{\mu (\alpha} g^{\beta) (\gamma} g^{\delta) \nu}).
\end{equation}
Here, $\{h_{\bf 1},  a_{\bf 1} \}, \{ h_{\bf 2}, a_{\bf 2} \} $ are two linearised solutions\footnote{A given linearised solution is specified 
by the collective perturbations $h_{\mu \nu}$ and $a_\mu$. We distinguish between the two 
solutions that enter as arguments of the currents by adding boldface indices ${\bf 1}$ and ${\bf 2}$.}, 
indices in parentheses are symmetrised and $^*$ indicates complex conjugation. 
Given the current, the symplectic flux through the boundary, ${\cal F}$, is by definition the pullback of the current to the surface $r=0$.
As usual, this is defined by evaluating the pullback at some cutoff surface $r = r_\epsilon$  and taking the limit $r_\epsilon \to 0$, so we write
\begin{equation} \label{flux}
{\cal F} = \lim_{r_\epsilon \to 0} \frac{i}{2} \int_{r = r_\epsilon} dt\, d^{d_s} x\, d \xi \sqrt{\gamma} n^\mu j_\mu ,
\end{equation}
where $n_\mu$ is the unit outward-pointing normal to the boundary.

We will also be interested in determining which linearised modes are normalizable, to determine which can be allowed to fluctuate in quantising the bulk theory. This requires us to define a suitable inner product. The inner product is usually defined in terms of the symplectic current by considering
\begin{equation}\label{ip}
 	( \{h_{\bf 1},  a_{\bf 1} \}, \{ h_{\bf 2}, a_{\bf 2} \}  ) = \frac{i}{2} \int_{\Sigma} \star j(\{h_{\bf 1},  a_{\bf 1} \}, \{ h_{\bf 2}, a_{\bf 2} \}),
\end{equation}
where $j$ is the symplectic current defined above, $\star$ is the Hodge dual and ${\Sigma} $ is a spacelike surface. However, in the Schr\"odinger background, there is no natural spacelike surface to consider; Schr\"odinger is not stably causal and therefore has no globally well-defined everywhere timelike vector field.

We will take $\Sigma$ to be a surface of constant $t$. We want to argue that this is a natural choice, as close as we can get to the usual construction in this case.  The irrotational condition \eqref{irrotcond} ensures that even the perturbed spacetime has a foliation, and asymptotically this foliation will be described by a constant $t$ surface.  Also, although this surface is null in the background spacetime, linear perturbations satisfying our boundary conditions will generically render it timelike.\footnote{The norm of the normal $n = dt$ is $n \cdot n = - r^4 H_{\xi\xi}$. The leading contribution to $H_{\xi\xi}$ comes from the source $\delta \hat e^+_\xi$ for $\mathcal E^\xi$, but this is set to zero by the boundary conditions. We will find in the linearised analysis below that the leading vev term is the particle number density, which is physically non-negative, so the normal becomes timelike unless the particle number density vanishes.} From an initial condition perspective, the fact that constant $t$ surfaces will continue to foliate the spacetime under perturbations makes them appealing.  Given that they are null in the background, our remaining concern would be information which propagates `parallel' to the constant $t$ surfaces; that is, along the $\xi$ direction.  However, we work in Fourier modes with constant $k_\xi$, and moreover focus on the sector with $k_\xi=0$ (even setting some $k_\xi \neq 0$ components to vanish via gauge choice). For the crude question we want to ask (for a given mode, is this inner product finite?) this seems sufficient. 

On surfaces of constant $t$, the expression \eqref{ip} simplifies to
\begin{equation}\label{ip 2}
 	( \{h_{\bf 1},  a_{\bf 1} \}, \{ h_{\bf 2}, a_{\bf 2} \}  ) = 
 	\frac{i}{2} \int_{\Sigma_t} dr d^{d_s} x d\xi \sqrt{g} j^t(\{h_{\bf 1},  a_{\bf 1} \}, \{ h_{\bf 2}, a_{\bf 2} \}  ).
\end{equation}

\section{Linearised analysis with spatial directions}
\label{spat}

We now specialise to the case with $d_s=2$.  (We will make some comments on differences for other values.) In this case, the analysis closely parallels the discussion for $z<2$ in \cite{Andrade:2014iia}. The main difference is that dependence on the null direction $\xi$ affects the linearised solutions at leading order, so that we need an independent discussion for the zero modes and the modes with non-zero $k_\xi$. The discussion of the zero modes is most interesting, both because this sector contains the stress energy complex of the dual field theory and because subtleties such as anomalies appear only in this sector.

As in \cite{Andrade:2014iia}, we are interested in identifying the modes corresponding to sources and vevs of dual operators. In many cases, this identification can be made simply using the scaling dimensions of the modes. Otherwise we use the flux to identify the vev as the mode canonically conjugate to the source following  \cite{Papadimitriou:2010as}. We consider first constant modes, independent of the boundary directions, 
to identify all sources and vevs, and then verify that they satisfy appropriate Ward identities in the non-constant cases.

We will discuss the case where the fields are independent of spatial coordinates $x^i$. The rotation symmetry in these directions is then unbroken, so we can decompose the linearised fields into a tensor, vector and scalar part with respect to this linearised symmetry. Below we will treat these initially for constant modes and then including dependence on $t, \xi$. To make this decomposition we should further decompose $H_{ij}$ into a trace and a trace free part, $H_{ij} = k \delta_{ij} + \bar H_{ij}$, where $\bar H_i^i = 0$.  The tensor mode is $\bar H_{ij}$. The vector modes are $H_{ti}$, $H_{\xi i}$ and $s_i$. The scalar modes are $H_{tt}$, $H_{t\xi}$, $H_{\xi\xi}$, $k$, $s_t$, $s_\xi$ and $s_r$ (the last is determined algebraically in terms of the other modes). We will always assume the $t,\xi$ dependence is harmonic, $e^{i \omega t + i k_\xi \xi}$, so in writing equations we will make the replacements $\partial_t \to i \omega$, $\partial_\xi \to i k_\xi$.

The extension to include dependence on the $x^i$ is a straightforward extension of the calculation for $z<2$ carried out in \cite{Andrade:2014iia}, so it is postponed to appendix \ref{app}.

\subsection{Tensor modes}

The tensor equation of motion is
\begin{equation} \label{tensoreq}
r^2 \bar H_{ij}'' - 3r \bar H_{ij}' - (  k_\xi ^2  + 2 k_\xi \omega r^2  ) \bar H_{ij} = 0.
\end{equation}
This is simpler than the equation in the $z<2$ case, so it can be solved in closed form for arbitrary $k_\xi, \omega$. For $k_\xi=0$, the solution is simply
\begin{equation}
\bar H_{ij} = \bar H_{ij}^{(0)} + \bar H_{ij}^{(4)} r^4,
\end{equation}
corresponding to the source and vev for the trace free part of the spatial stress tensor $\Pi_{ij}$. For non-zero $k_\xi$, the solution is
\begin{equation}
	\bar H_{ij} = \bar H_{ij}^{(-)}  r^2 J_\nu ( - i \sqrt{2 k_\xi \omega} r ) +  \bar H_{ij}^{(+)} r^2 Y_\nu ( - i \sqrt{2 k_\xi \omega} r ),
\end{equation}
\noindent where $J_\nu$ and $Y_\nu$ are Bessel functions of the first and second kind and $\nu = \sqrt{4 + k_\xi^2}$. These modes are 
the source and vev for some tensor operator; the asymptotics $r^{2 \pm \nu}$ tell us that this is an operator of dimension $2+\nu$. This is an irrelevant operator for all $k_\xi >0$.

\subsection{Vector modes}

The vector equations of motion are
\begin{align}
\label{vi xi kxi omega}
	r^2 s_i'' - 3 r s_i' - [5 + k_\xi^2 + 2 k_\xi \omega r^2 ]  s_i + 2 r H_{\xi i}' + 2 H_{\xi i} &= 0, \\	
\label{Hti xi kxi omega}
	k_\xi [  r ( H_{\xi i}' + H_{t i}' ) + ( H_{\xi i} - H_{t i} - 2  s_i)   ] + \omega r^{2} [ r H_{\xi i}' + H_{\xi i}  ] &= 0 , \\
\label{Hxii xi kxi omega}
	r^2 H_{\xi i}'' - r H_{\xi i}' - (3+ r^2 k_\xi \omega) H_{\xi i} + k_\xi^2 H_{ti} & = 0,
\end{align}
and
\begin{align}
\nonumber
	r^2 H_{ti}'' & -  5r  H_{ti}' + [5 - k_\xi^2 - r^2 k_\xi \omega] H_{ti} \\
\label{2nd order Hti kxi omega}
	& + 2 [ 5 s_i - H_{\xi i} - r (s_i + H_{\xi i})' ] + (r^2 k_\xi \omega + r^{4} \omega^2) H_{\xi i}= 0 .
\end{align}

\subsubsection{Zero modes}

For $k_\xi = \omega = 0$, \eqref{Hti xi kxi omega} is trivially satisfied, and we solve (\ref{vi xi kxi omega},\ref{Hxii xi kxi omega},\ref{2nd order Hti kxi omega}). The solutions are obtained as the limit as $z \to 2$ of the solutions in \cite{Andrade:2014iia}:
\begin{align}
	s_i &= s_i^{(-)} r^{-1} + H_{\xi i}^{(+)} r^3 + s_i^{(+)} r^5,  \\
	H_{ti} &= -s_i^{(-)} r^{-1} + H_{ti}^{(-)} r - H_{\xi i}^{(+)} r^3 + H_{ti}^{(+)} r^5 , \\
	H_{\xi i} &= (H_{\xi i}^{(-)}+ s_i^{(-)}) r^{-1} + H_{\xi i}^{(+)} r^3 .
\end{align}
We have chosen to define and normalise the independent modes so that the solutions with a $(-)$ superscript correspond to the sources, coming from the constant modes in the frame fields: $s_i^{(-)}$ is the source term for the energy flux $\mathcal E^i$, $ H_{\xi i}^{(-)}$ is the source term for the extra vector operator $\mathcal P^\xi_i$, and $H_{t i}^{(-)}$ is the source term for the momentum density $\mathcal P_i = \rho^i$. The modes with a $(+)$ superscript should then correspond to the vevs of these operators. By dimensions alone we see that $\langle \mathcal P_i \rangle \sim H_{\xi i}^{(+)}$. The vevs $\langle \mathcal E^i \rangle$ and $\langle \mathcal P^\xi_i \rangle$ should be related to $H_{t i}^{(+)}$ and $s_i^{(+)}$.

For $k_\xi =0$, the flux can also be smoothly obtained as the limit $z \to 2$ of the results in \cite{Andrade:2014iia}:
\begin{align}
\nonumber
 	 {\cal F} =  - i \int_{r = 0} dt\, d^{2} x \, d \xi \, & \bigg [  H_{\xi i}^{(-)} \wedge (2 H_{t i}^{(+)} 
 	 - s_i^{(+)}) + 2 H_{t i}^{(-)} \wedge H_{\xi i}^{(+)}    \\
 	 & + s_i^{(-)} \wedge (2 H_{t i}^{(+)} +  2 s_i^{(+)}) \bigg] ,
\end{align}
\noindent where  $A\wedge B=A_{\bf 1} B_{\bf 2}-A_{\bf 2} B_{\bf 1}$, with $\bf 1, \bf 2$ labelling the two linearised solutions 
which define the current. This enables us to identify, up to an overall normalization which we neglect for simplicity,
\begin{equation}
\langle \mathcal P_i \rangle = 2 H_{\xi i}^{(+)}, \quad  \langle \mathcal P_i^\xi \rangle = 2 H_{ti}^{(+)} -  s_i^{(+)}, \quad \langle \mathcal E^i \rangle = 2 H_{ti}^{(+)} + 2 s_i^{(+)}.
\end{equation}

For non-zero $\omega$, the solution is modified first in that \eqref{Hti xi kxi omega} is no longer trivially satisfied; it sets $H_{\xi i}^{(+)} = 0$, corresponding to the expected Ward identity $\partial_t \mathcal P_i = 0$. Secondly, there is an $\omega^2$ term in \eqref{2nd order Hti kxi omega}, which implies the solution is modified by subleading contributions. In this case there is a single subleading contribution. The solution is
\begin{align}
	s_i &= s_i^{(-)} r^{-1}  + s_i^{(+)} r^5,  \\
	H_{ti} &= -s_i^{(-)} r^{-1} + H_{ti}^{(-)} r +\frac{1}{4} \omega^2 H_{\xi i}^{(-)}  r^3 + H_{ti}^{(+)} r^5,  \\
	H_{\xi i} &= H_{\xi i}^{(-)} r^{-1}  .
\end{align}

\subsubsection{Non-zero $k_\xi$}

For non-zero $k_\xi$, the leading-order solution with $\omega =0$ can be written as
\begin{align}
	H_{\xi i} &=  H_{\xi i}^{(diff)} r^{-1} + r^{1-\nu}  H_{\xi i}^{(-)}  + r^{1+\nu}  H_{\xi i}^{(+)}
	+   (\nu - 2 )  k_\xi^2  H_{\xi i}^{(3-)} r^{3 - \nu}
	+  (\nu + 2 )  k_\xi^2  H_{\xi i}^{(3+)} r^{3 + \nu},  \\
\nonumber
s_i &= r^{1-\nu}  H_{\xi i}^{(-)}  + r^{1+\nu}  H_{\xi i}^{(+)} \\
	&- r^{3 - \nu} H_{\xi i}^{(3-)}  [24 (\nu - 2)  + k_\xi^2 (\nu - 8 )  ]
	- r^{3 + \nu} H_{\xi i}^{(3+)}  [24 (\nu + 2)  + k_\xi^2 (\nu + 8 )  ] ,  \\
\nonumber
H_{t i} &= - r^{1-\nu}  H_{\xi i}^{(-)}  - r^{1+\nu}  H_{\xi i}^{(+)} \\
	&- r^{3 - \nu} H_{\xi i}^{(3-)}  [12 (\nu - 2)  + k_\xi^2 (\nu - 6 )  ]
	- r^{3 + \nu} H_{\xi i}^{(3+)}  [12 (\nu + 2)  + k_\xi^2 (\nu + 6 )  ],
\end{align}
where $\nu$ is as before. We see that we have a source and a vev for an operator of dimension $1+\nu$, and an operator of dimension $3+\nu$. The latter is irrelevant for all $k_\xi >0$, the former is relevant for $k_\xi^2 < 5$. The mode $H_{\xi i}^{(diff)}$ whose dimension is independent of $k_\xi$ is a pure diffeomorphism. We see that this mode corresponds to a non-zero mode of $\delta \hat e^I_\xi$, as expected. As argued in section \ref{kk}, the natural approach is to gauge-fix these $\xi$-dependent diffeomorphisms by setting this mode to zero. The physical content in this non-zero mode sector is thus a pair of vector operators of dimensions $1+\nu$, $3+\nu$. Including non-zero $\omega$ will lead to an infinite series of subleading corrections in powers of $k_\xi \omega r^2$.

\subsection{Scalar modes}

We consider now the scalar modes $H_{tt}$, $H_{t \xi}$, $H_{\xi \xi}$, $k$, $s_r$, $s_t$, $s_\xi$. They are governed by the equations
\begin{align}
0 =&\, \left(k_\xi + 2\omega  r^{2}\right) H_{\xi \xi}   - 2k_\xi  k + k_\xi r s_t' -  2k_\xi s_t
\label{eq sc 1 kxi w}
 +( k_\xi  + \omega  r^{2} ) r s_\xi' - i \left( k_\xi^2 + 2  k_\xi \omega  r^2 + 8 \right) s_r ,
\\\nonumber
	0=&\, -2 r H_{\xi \xi}' + \left(\frac{1}{2} \omega^2 r^{4} - 2  \right) H_{\xi \xi}
	- 3 r H_{t\xi}' - r^2 k_\xi \omega H_{t \xi}  + \frac{1}{2} k_\xi ^2 H_{tt}
\\\label{eq sc 4 kxi w}
	&\, -3 r k' -  \left(k_\xi ^2  + 2 k_\xi \omega r^2  \right) k
	+ [ r s_\xi' + 4 s_\xi -i k_\xi  s_r ],
\\\nonumber
0=&\,	\frac{\omega}{2} r H_{\xi \xi}' + \left(\frac{1}{2} k_\xi  r^{-2}+\frac{3}{2} \omega  \right) H_{\xi \xi}  +
	\left(\frac{\omega }{2}-\frac{1}{2} k_\xi r^{-2} \right) r H_{t \xi}'  \\
&\,	- \frac{1}{2} k_\xi r^{-2} r H_{tt}'  + \omega r k'  +  r^{-2} k_\xi  [  H_{tt}  -  k +  s_t]
\label{eq sc 5 kxi w}	 - \omega s_\xi - 4i  r^{-2} s_r,
\\
\label{eq sc 6 kxi w}
0=&\,	-\frac{1}{2} \omega r^{2} r H_{\xi \xi} ' -  \left(\omega   r^{2} + \frac{k_\xi}{2}\right) H_{\xi \xi}
	+\frac{1}{2} k_\xi r H_{t \xi} ' + k_\xi r k' ,
\\
\nonumber
0=&\,	\frac{1}{2} r^2 H_{\xi \xi}''  + r^2 H_{t \xi} '' + \frac{1}{2} r^2 H_{tt}''  + r^2 k''  - \frac{1}{2} r  H_{\xi \xi}' - 4 r H_{t \xi} ' - \frac{7}{2}  r H_{tt}' + 6 H_{tt} -2 r k'
\\
&\, - \left( k_\xi^2 + 2 k_\xi \omega r^2  + \omega^2 r^{4} \right) k - 2  r s_t' + 12 s_t
\label{eq sc 7 kxi w}- r s_\xi' + 4  s_\xi + i  \left( k_\xi  + 2 \omega r^2 \right) s_r ,
\\\label{eq sc 9 kxi w}
0=&\,	r^2 H_{\xi \xi}'' + r H_{\xi \xi}' - 4  H_{\xi \xi} - 2 k_\xi^2  k  ,
\\\nonumber
0=&\,
	r^2 H_{\xi \xi} '' + 2 r^2 H_{t \xi}'' + r^2 k''  + \left(\omega ^2 r^{4}-4\right) H_{\xi \xi}  - 6 r H_{t \xi}'
	+ k_\xi^2 H_{tt}
\\
&\, -2 r H_{\xi \xi}'  - 3 r k' - \left(k_\xi^2 + 2 k_\xi \omega r^2  \right) k
\label{eq sc 10 kxi w}	- 2 k_\xi \omega r^2  H_{t \xi}  + 2  [  i k_\xi  s_r  -  r s_\xi' + 4 s_\xi] .
\end{align}

In addition, we have the radial-component gravitational constraint equations:
\begin{align}
0=&\,	r^2 s_\xi ''  +  4H_{\xi \xi}  + k_\xi^2  s_t +  3 r s_\xi ' + 2 r H_{\xi \xi} '
\label{eq sc 2 kxi w}	- \left(k_\xi \omega r^2  + 8  \right) s_\xi  + i  k_\xi   (2 s_r  -  r s_r'),
\\\nonumber
0=&\,	r^2 s_\xi ''  +  r^2 s_t ''  - k_\xi \omega r^2 s_t
	+  \left(\omega ^2 r^{4}-8\right)  s_\xi   -   3 r  s_\xi'  - 5 r s_t'
\\\label{eq sc 3 kxi w}
&\,	+	 r H_{\xi \xi} '  -  2 r k'  +  2 i \left( 2k_\xi   +  \omega  r^2  \right) s_r
	- i  \left( k_\xi  + \omega  r^{2} \right) r s_r' ,
\\
0=&\,	\frac{1}{2} r^2 H_{\xi \xi} '' + \frac{1}{2} r^2 H_{t \xi} '' + r^2 k'' \label{eq sc 8 kxi w}
	-\frac{3}{2} r H_{t \xi} '   - 3 r k'  - \left( k_\xi^2 + k_\xi \omega r^2 \right) k  .
\end{align}

\subsubsection{Zero modes}

For $k_\xi = \omega = 0$, (\ref{eq sc 1 kxi w},\ref{eq sc 5 kxi w},\ref{eq sc 6 kxi w}) are automatically satisfied if $s_r=0$, and (\ref{eq sc 2 kxi w},\ref{eq sc 3 kxi w},\ref{eq sc 8 kxi w}) are non-trivial equations. The solution for the scalar modes is not a smooth limit of the solution in  \cite{Andrade:2014iia}, as that solution involved factors of $(z-2)^{-1}$, so it does not have a smooth limit. This arises because some powers of $r$ in the mode solution which are distinct for $z \neq 2$ coincide for $z=2$. As a result the solution involves logarithms. The solution is
\begin{align}
	s_r &= 0, \\
	H_{\xi \xi} &= 2H_{\xi \xi}^{(-)} r^{-2} + H_{\xi \xi}^{(+)} r^{2}, \\
	s_\xi &= r^{-2} (s_\xi ^{(-)}+H_{\xi\xi}^{(-)})  + H_{\xi \xi}^{(+)} r^2 + s_\xi ^{(4)} r^4, \\
	k &= \frac{1}{3}  s_\xi^{(-)} r^{-2} + 2k^{(0)} + \frac{1}{2} H_{\xi \xi}^{(+)} r^2 + k^{(4)} r^4, \\
	H_{t \xi} &= - r^{-2} ( H_{\xi \xi}^{(-)} + \frac{2}{3} s_\xi^{(-)} ) + s_t^{(0)} + H_{t \xi}^{(0)} - \frac{1}{2} 	H_{\xi \xi}^{(+)} r^2
	+  (\frac{2}{3} s_\xi^{(4)} -  k^{(4)} ) r^4, \\
	s_t &= -\frac{1}{3} s_\xi^{(-)}  r^{-2} + s_t^{(0)} - \frac{3}{2} H_{\xi \xi}^{(+)} r^2
	- (k^{(4)} + s_\xi^{(4)}) r^4 + s_t^{(+)} r^6, \\
\nonumber
H_{tt} &= \frac{1}{3}  s_\xi^{(-)}  r^{-2} -2 s_t^{(0)} - H_{\xi \xi}^{(+)} r^2 (1 + 4 \log  r) + 2 H_{tt}^{(-)} r^2 \\
	&+ \left(  2 k^{(4)} - \frac{10}{3} s_\xi^{(4)} \right) r^4 + H_{tt}^{(+)} r^6. \label{Htt}
\end{align}

In the familiar AdS case, there are logarithmic terms in the Fefferman-Graham expansion for even boundary dimension, which are related to the anomaly in the scaling symmetry \cite{Henningson:1998gx}.\footnote{As noted in \cite{deHaro:2000xn}, the log terms in the FG expansion are proportional to the variation of the integrated anomaly.} One might expect that the logarithm appearing in \eqref{Htt} would similarly contribute to an anomaly in the anisotropic scaling symmetry here. However, the anomaly is determined by the variation of the action under the scaling symmetry $\delta_\sigma S$, and because the background metric has $g^{tt} = 0$, $H_{tt}$ cannot contribute to the action at linear order. Thus there is no anomaly term coming from \eqref{Htt}. The logarithmic term will however have implications for boundary conditions, as in the case of a scalar field at the BF bound, see e.g. \cite{Witten:2001ua}. The appearance of the logarithm implies that the only scale-invariant boundary condition is one which fixes the coefficient of the logarithm $H_{\xi\xi}^{(+)}$. If we impose a boundary condition fixing $H_{tt}^{(-)}$ at some scale, this evolves into a mixed boundary condition under scale transformations: $r \to \lambda r$ maps $H_{tt}^{(-)} \to \lambda^{-2} (H_{tt}^{(-)} + 2 \log \lambda H_{\xi\xi}^{(+)})$. This is surprising because $H_{\xi\xi}^{(+)}$ corresponds to a component of the stress energy complex, whereas we had been assuming that our boundary conditions would fix the boundary geometry, encoded in the $(-)$ modes. We will discuss these issues below after considering the flux and inner product. 

We calculate the flux to identify the canonically conjugate pairs of modes and hence identify the components of the stress tensor complex. The flux is a smooth limit of the expression for $z <2$ in  \cite{Andrade:2014iia},
\begin{align}
\nonumber 
	{\cal F} = i \int_{r = 0} dt\, d^{d_s} x \, d \xi \, & \bigg [
	 s_t^{(0)} \wedge \left (2 k^{(4)} - \frac{5}{3} s_\xi^{(4)} \right ) +  H_{t\xi}^{(0)} \wedge \left (2 k^{(4)} + \frac{4}{3}  s_\xi^{(4)}  \right) \\
\nonumber
	&\,+ k^{(0)} \wedge \left ( - 4 k^{(4)} + \frac{10}{3}  s_\xi^{(4)} \right) - 2  H_{tt}^{(-)} \wedge H_{\xi\xi}^{(+)} \\
	&\,-  H_{\xi\xi}^{(-)} \wedge \left( 2 H_{tt}^{(+)} + s_t^{(+)} \right) - 3 s_\xi^{(-)} \wedge s_t^{(+)} \bigg ] . \label{sflux}
\end{align}
This implies the identifications
\begin{equation}\label{E 1}
\langle \mathcal E \rangle = - 2 k^{(4)} + \frac{5}{3}  s_\xi^{(4)},
\end{equation}
and
\begin{equation}\label{Pii 1} 
\langle \Pi_i^i \rangle  = \langle \Pi_1^1 \rangle + \langle \Pi_2^2 \rangle =  4 k^{(4)} - \frac{10}{3}  s_\xi^{(4)}.
\end{equation}
The Ward identity from the scaling invariance is 
\begin{equation}\label{trace ward 1}
	2 \mathcal E + \Pi_i^i  =0
\end{equation}
\noindent which is indeed satisfied by these vevs. This confirms that, as argued above, the scale anomaly vanishes for these modes despite the presence of logarithms in the radial profiles. 
The other vevs are
\begin{equation}
\langle \mathcal E^\xi \rangle = 2 H_{tt}^{(+)} +  s_t^{(+)},
\end{equation}
\begin{equation}
\langle \rho \rangle = 2 H_{\xi\xi}^{(+)}, \quad \langle \rho^\xi \rangle = -2 k^{(4)} - \frac{4}{3}  s_\xi^{(4)} ,
\end{equation}
and
\begin{equation}
\langle O \rangle =  3 s_t^{(+)}.
\end{equation}

We would now like to consider the possible boundary conditions. To make the flux through the boundary vanish, boundary conditions should fix one of each conjugate pair in \eqref{sflux}. In addition, we want to fix the non-normalizable modes for which the inner product \eqref{ip} diverges. For $k_\xi = 0$, the inner product is finite in the UV provided we set $H_{\xi\xi}^{(-)} = s_\xi^{(-)} = 0$. If we allow for non-zero $k_\xi$, to cancel subleading divergences we must set also $k^{(0)} = 0$. With the conditions $H_{\xi\xi}^{(-)} = s_\xi^{(-)} = k^{(0)}= 0$, the only divergence that is left is proportional to $ k_\xi r^{-1} (H_{\xi \xi}^{(+)} H_{t \xi}^{(0)})$, so we need at least one of $H_{t \xi}^{(0)} = 0$ or $H_{\xi \xi}^{(+)} = 0 $. A consistent choice is then to set $H_{\xi\xi}^{(-)} = s_\xi^{(-)} =  s_t^{(0)} = k^{(0)}= H_{t \xi}^{(0)} = 0$; then both $H_{tt}^{(-)}$ and $H_{\xi\xi}^{(+)}$ are normalizable, and we can choose to fix either of them. We would originally have thought we wanted to fix $H_{tt}^{(-)}$, corresponding to fixing the boundary geometry, but as noted above, this is not a scale invariant boundary condition; the only scale invariant boundary condition is to fix instead $H_{\xi\xi}^{(+)} = 0$. In the field theory, this is fixing the particle number density, rather than its source. 

For non-zero $\omega$, (\ref{eq sc 1 kxi w},\ref{eq sc 5 kxi w},\ref{eq sc 6 kxi w}) become non-trivial. One of these fixes $s_r$; the other two linear combinations give the expected Ward identities $\partial_t \mathcal E = 0$ and $\partial_t \rho =0$, setting $H_{\xi\xi}^{(+)} = 0$ and $k^{(4)} = \frac{5}{6} s_\xi^{(4)}$. This confirms the identification of the modes corresponding to components of the stress tensor, which is summarised in table 1.

\begin{table}
\begin{center}
\begin{tabular}{|l||l|l|}
\hline
Operator & Source & Expectation value \\
\hline
$\mathcal E$ & $\delta \hat e^+_t = s_t^{(0)}$ & $- 2 k^{(4)} + \frac{5}{3}  s_\xi^{(4)}$ \\
$\mathcal E^i$ & $\delta \hat e^+_i = s_i^{(-)}$ & $2 H_{ti}^{(+)} + 2 s_i^{(+)}$ \\
$\mathcal E^\xi$ & $\delta \hat e^+_\xi = H_{\xi\xi}^{(-)}$ & $2 H_{tt}^{(+)} +  s_t^{(+)}$ \\
$\rho$ & $\delta \hat e^-_t = H_{tt}^{(-)}$ & $2 H_{\xi\xi}^{(+)}$ \\
$\mathcal P_i = \rho^i$ & $\delta \hat e^-_i = H_{ti}^{(-)}$ & $2 H_{\xi i}^{(+)}$ \\
$\rho^\xi$ & $\delta \hat e^-_\xi = H_{t\xi}^{(0)}$ & $-2 k^{(4)} - \frac{4}{3}  s_\xi^{(4)}$ \\
$\Pi^1_1 + \Pi^2_2$ & $\delta \hat e^I_i = k^{(0)}$ & $4 k^{(4)} - \frac{10}{3} s_\xi^{(4)}$ \\
$\Pi^1_1 - \Pi^2_2$, $\Pi^1_2$ & $\delta \hat e^I_j =  \bar H_{ij}^{(0)}$ & $\bar H_{ij}^{(4)}$ \\
$\mathcal P^\xi_i$ & $\delta \hat e^I_\xi  = H_{\xi i}^{(-)}$ & $2 H_{ti}^{(+)} - s_i^{(+)}$ \\
$O$ & $s_\xi^{(-)}$ & $3 s_t^{(+)}$ \\
\hline
\end{tabular}
\end{center}
\caption{The identification of linearised modes with sources and vevs for the operators in the dual field theory. Note that the sources are the zero modes of the indicated frame field component.}
\end{table}

 There are also subleading terms in $\omega^2$ in solving the other equations. As in the vector case, there are a finite number of subleading terms, and the full solution for non-zero $\omega$ is
\begin{align}
	s_r &=  - i \frac{\omega}{2} \left( \frac{1}{2} (H_{\xi \xi}^{(-)} - s_\xi^{(-)} ) + s_\xi^{(4)} r^6 \right), \\
	H_{\xi \xi} &= 2H_{\xi \xi}^{(-)} r^{-2}  ,\\
	s_\xi &= r^{-2} (s_\xi ^{(-)}+H_{\xi\xi}^{(-)})  + s_\xi ^{(4)} r^4 ,\\
	k &= \frac{1}{3}  s_\xi^{(-)} r^{-2} + 2k^{(0)}   - \frac{1}{6} H_{\xi \xi}^{(-)} \omega^2 r^2 + \frac{5}{6} s_\xi^{(4)} r^4,  \\
	H_{t \xi} &= - r^{-2} ( H_{\xi \xi}^{(-)} + \frac{2}{3} s_\xi^{(-)} ) + s_t^{(0)} + H_{t \xi}^{(0)}   - \frac{1}{6} s_\xi^{(+)} r^4 + \frac{1}{3} H_{\xi \xi}^{(-)} \omega^2 r^2,\\
	s_t &= -\frac{1}{3} s_\xi^{(-)}  r^{-2} + s_t^{(0)}
	- \frac{11}{6} s_\xi^{(4)} r^4 + s_t^{(+)} r^6  + \frac{1}{48} ( 5 H_{\xi \xi}^{(-)} + 3 s_\xi^{(-)}  ) \omega^2 r^2  + \frac{3}{48}  \omega^2 r^8 s_\xi^{(+)},
\\ \nonumber
H_{tt} &= \frac{1}{3}  s_\xi^{(-)}  r^{-2} -2 s_t^{(0)}  + 2 H_{tt}^{(-)} r^2 - \frac{5}{3} s_\xi^{(4)}  r^4 + H_{tt}^{(+)} r^6 + \frac{1}{24} (2 H_{\xi \xi}^{(-)} - s_\xi^{(-)} ) \omega^2 (1 + 4 \log r)  r^2
\\
	& -  \frac{1}{2} k^{(0)} \omega^2 r^4 +  \frac{1}{96} H_{\xi \xi}^{(-)} \omega^4 (1 - 4 \log r)  r^6
	+ \frac{1}{72} \omega^2 s_\xi^{(+)} r^8.
\end{align}
We see that since the Ward identity sets $H_{\xi\xi}^{(+)} = 0$, the previous logarithmic term is absent for non-zero $\omega$, but there are new derivative terms with logarithms. As before, they will not give an anomaly for the scaling symmetry, as $H_{tt}$ cannot contribute to the action. These now involve the $(-)$ modes, which we are used to thinking of as sources, so it is not worrying that the scale invariant boundary condition is to fix these modes.  

If we study the scalar system for $d_s=1$ or $d_s=3$, the logarithmic term involving $H_{\xi\xi}^{(+)}$ in the constant modes is absent, but this logarithmic term in the $\omega$-dependent modes persists.

\subsubsection{Non-zero $k_\xi$}

For non-zero $k_\xi$, the leading-order solution with $\omega=0$ has three independent bulk diffeomorphism modes with dimensions which are independent of $k_\xi$, and six modes whose dimensions depend on $k_\xi$. The bulk diffeomorphism modes are generated by 
\begin{equation}
	\chi = r \chi^r_0 \partial_r + \left( \chi^t_0 - \frac{i}{2} k_\xi r^2 \chi^r_0 \right) \partial_t 
	+ ( \chi^\xi - i k_\xi \chi^r_0 \log r  ) \partial_\xi.
\end{equation}
The resulting modes are
\begin{align}
\label{scalar soln diffeo1}
s_r &=  - i k_\xi \chi^r_0, \\
s_t &=  - 2 \chi^r_0,  \\
s_\xi &=  i k_\xi \chi^t_0 r^{-2} + \frac{1}{2} k_\xi^2 \chi^r_0, \\
H_{tt} &=  4 \chi^r_0, \\
H_{t \xi} &=   - i k_\xi \chi^t_0 r^{-2}  - \frac{1}{2}(4 + k_\xi^2) \chi^r_0 + i k_\xi \chi^\xi_0 + k_\xi^2 \log r \chi^r_0, \\
H_{\xi \xi} &= 2 i k_\xi \chi^t_0 r^{-2} + k_\xi^2 \chi^r_0, \\
\label{scalar soln diffeo2}
k &=  - \chi^r_0.
\end{align}
As argued in section \ref{kk}, we can gauge-fix the boundary diffeomorphism symmetry by setting $\delta \hat e^+_\xi$ and $\delta \hat e^-_\xi$ to zero; this corresponds to setting $\chi^t_0$ and $\chi^\xi_0$ to zero.  We see that there is a logarithmic term involving $\chi^r_0$; as this is now in $H_{t\xi}$, it may contribute to the anomaly. 
However, we see from \eqref{trace ward 1}, \eqref{Pii 1} \eqref{E 1} than only contributions to $k^{(4)}$ and $s_\xi^{(4)}$ 
participate in the corresponding Ward identity, and, because these coefficients are zero in the solution 
\eqref{scalar soln diffeo1}-\eqref{scalar soln diffeo2}, the anomaly must vanish. 

\begin{figure}[h]
\begin{center}
\includegraphics[scale=0.5]{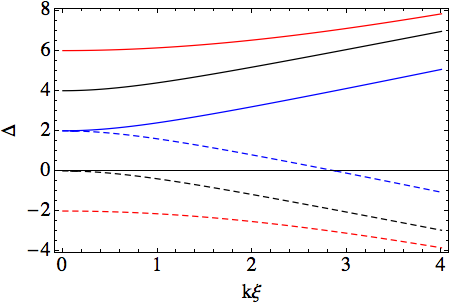}
\caption{Scaling dimensions of the non-diffeo modes in the scalar sector for $d_s= 2$, $z =2$ as a function of $k_\xi$.
Pairs of dimensions that add to four are plotted in the same color, using solid and dashed lines.}
\label{deltas}
\end{center}
\end{figure}

The physical degrees of freedom in the non-zero momentum sector are in the other six modes. The scaling of the fields for these modes is $r^\Delta$ where $\Delta$ satisfies a sixth-order equation
\begin{align}\label{eq Delta}
\nonumber
	& 3 \Delta^6 - 36 \Delta^5 + (120 - 9 k_\xi^2) \Delta^4 + 72 k_\xi^2 \Delta^3 + \\
	& (9 k_\xi^4 - 112 k_\xi ^2 - 528) \Delta^2	- 4 (9 k_\xi^4 + 32 k_\xi^2 - 144) - (3 k_\xi^6 + 8 k_\xi^4 - 272 k_\xi^2) = 0.
\end{align}
The solutions to \eqref{eq Delta} can be found in closed from, although the explicit expressions are not very illuminating. They are plotted in figure \ref{deltas}. The solutions come in pairs which sum to $4$, so we can identify them as the sources and vevs for three scalar operators. Two of these operators are irrelevant for all $k_\xi >0$, and the third becomes irrelevant at some critical value of $k_\xi$, as we can see from figure \ref{deltas}.

\section{Linearised solutions for $d_s=0$}
\label{dszero}

We now consider the linearised solutions for $d_s=0$. This case will clearly have a different behaviour from our present perspective: the absence of spatial directions modifies the structure of the linearised equations, and implies that the stress energy complex simplifies, as there are now no spatial fluxes. This case has also been extensively studied in previous work \cite{Guica:2010sw,Costa:2010cn,Guica:2011ia,vanRees:2012cw}, under the name of null warped AdS$_3$, so we consider it in detail to make contact between our analysis and previous work. In this case, we only have the analogue of the scalar modes in the above discussion.  However, the Ward identities in this case become degenerate, so new behavior arises which was not present in the $d_s=2$ case. 

The equations of motion read
\begin{align}
\label{eq1 scalar z=2 D=3}
	0 &= ( k_\xi +2 r^2 \omega) H_{\xi \xi} - 2 k_\xi s_t - i (4+ k_\xi^2 + 2 \omega k_\xi r^2 ) s_r +  ( k_\xi +r^2 \omega ) r s_\xi '
	+ k_\xi r s_t', \\
\nonumber
0 &= 2 k_\xi H_{tt} + 2 k_\xi s_t - 2 r^2 \omega s_\xi - 4 i s_r+ ( k_\xi+ 3 r^2 \omega ) H_{\xi \xi} \\
\label{eq2 scalar z=2 D=3}
 & + \omega r^3 H_{\xi \xi}'  + (r^2 \omega - k_\xi ) r H_{t \xi} ' -  k_\xi r H_{tt}' , \\
 \label{eq3 scalar z=2 D=3}
0 &= \omega r^3 H_{\xi \xi} ' +  ( k_\xi + 2 r^2 \omega ) H_{\xi \xi}  - k_\xi  r H_{t \xi}' ,\\
\label{eq4 scalar z=2 D=3}
0 &= 4 s_\xi + H_{\xi \xi} (r^4 \omega^2 - 2) - 2 k_\xi r^2 \omega H_{t \xi} + 
k_\xi^2 H_{tt} - 2 i k_\xi s_r -2 r H_{\xi \xi}'   - 2 r H_{t \xi} ' + 2 r s_\xi' ,  \\
\nonumber
0 &=  16 s_t + 2 H_{\xi \xi} + 4 s_\xi + r H_{\xi \xi}'   - 4 r H_{t \xi} '  
- 5 r H_{tt}' + 8 H_{tt}  + 2 i s_r ( k_\xi + 2 r^2\omega ) \\ 
\label{eq5 scalar z=2 D=3}
 &- 2 r s_\xi'  - 4 r s_t' + 2 r^2 H_{t \xi} '' + r^2 H_{\xi \xi} '' + r^2 H_{tt}''  ,\\
\label{eq6 scalar z=2 D=3}
0 &=  3 r H_{\xi \xi} ' + r^2 H_{\xi \xi} '' , \\
\label{eq7 scalar z=2 D=3}
	0 &= 4 H_{\xi \xi} + k_\xi^2 s_t - (4 + \omega k_\xi r^2 ) s_\xi  + 2 r H_{\xi \xi}'  - i k_\xi r s_r' + r s_\xi' + r^2 s_\xi'' ,\\
\nonumber
	0 & =  2 i k_\xi s_r - k_\xi r^2 \omega s_t   + 
	 ( r^4 \omega^2 - 4 ) s_\xi  + r H_{\xi \xi}'  -  i ( k_\xi + r^2 \omega ) r s_r'  \\ 
\label{eq8 scalar z=2 D=3}
	 &- r s_\xi' - 3 r s_t' +  r^2 s_\xi'' + r^2 s_t''  .
\end{align}
For constant modes, \eqref{eq1 scalar z=2 D=3}, \eqref{eq2 scalar z=2 D=3}, \eqref{eq3 scalar z=2 D=3} trivialise, becoming simply
$s_r = 0$. A full solution can be found by solving  \eqref{eq4 scalar z=2 D=3}-\eqref{eq8 scalar z=2 D=3}. For general $k_\xi$, $\omega$, 
a complete set of equations is given by \eqref{eq1 scalar z=2 D=3}-\eqref{eq6 scalar z=2 D=3}, and we can easily check that  
\eqref{eq7 scalar z=2 D=3} and \eqref{eq8 scalar z=2 D=3} follow as a consequence of them.

\subsection{Zero modes}

The solution for constant  modes is again not a smooth limit of the one in  \cite{Andrade:2014iia} because of the appearance of factors of $(z-2)^{-1}$, which are replaced by logarithmic terms. The solution is
\begin{align}
\label{Hxixi constant modes D= 3 z=2}
	H_{\xi \xi} &=   2H_{\xi \xi}^{(-)} r^{-2}  +  H_{\xi \xi}^{(+)} ,\\
	H_{t \xi} &=  -H_{\xi \xi}^{(-)}  r^{-2}  + s_t^{(0)} +  H_{t \xi}^{(0)}  + H_{\xi \xi}^{(+)}(1+ \log r)+ 2  s_\xi^{(+)} r^2 ,\\
	H_{tt} &=  - 2 s_t^{(0)} + H_{\xi \xi}^{(+)} (1+ 2\log r)  +
	2 H_{tt}^{(-)}   r^2  - 4  s_\xi^{(+)} (1+2 \log r) r^2  + H_{tt}^{(+)} r^4  , \\
	s_r &= 0 ,\\
	s_t &=  -\frac{1}{3} s_\xi^{(-)} r^{-2} + s_t^{(0)} - H_{\xi \xi}^{(+)} \log r -  s_\xi^{(+)} r^2 + s_t^{(+)} r^4 , \\
\label{sxi constant modes D= 3 z=2}
	s_\xi &=  (s_\xi^{(-)} + H_{\xi\xi}^{(-)}) r^{-2} + H_{\xi \xi}^{(+)}   +  s_\xi^{(+)} r^2.
\end{align}
As in the higher-dimensional case, we have logarithms. The logarithms in $H_{tt}$ and $s_t$ cannot contribute to the anomaly at linear order for the same reason as above: since the background has $g^{tt} = 0$, we can't build a scalar out of $H_{tt}$ or $s_t$. The logarithm in $H_{t\xi}$ could in principle contribute to the anomaly, but this is not the case as we shall see below. 
The logarithms lead to inhomogeneous transformations under scaling, as in $d_s=2$. When $r \to \lambda r$, $s_t^{(0)} \to s_t^{(0)} - H_{\xi\xi}^{(+)} \log \lambda$, and $H_{tt}^{(-)} \to H_{tt}^{(-)} + 4 s_\xi^{(+)} \log \lambda$. These  again have the surprising feature that modes that correspond to the components of the boundary geometry have an inhomogeneous transformation depending on vev modes, implying that we can't have a scale invariant boundary condition that fixes the boundary geometry. 

To determine the identification of the components of the stress energy complex, we calculate the flux for these constant modes. This is again the $z \to 2$ limit of the previous expression (for the appropriate definition of the modes),
\begin{equation}
     {\cal F} = -i \int_{r = 0} dt \, d \xi\,  \bigg[  H_{\xi \xi}^{(-)} \wedge  H_{tt}^{(+)} + H_{tt}^{(-)}\wedge H_{\xi \xi}^{(+)}  + 2 s_\xi^{(-)} \wedge  s_t^{(+)}  - 2H_{t \xi}^{(0)}\wedge s_\xi^{(+)}  \bigg] .
\end{equation}
Note we have redefined the modes here relative to the $z<2$ case in order to cleanly separate the vevs from the sources.
This enables us to identify the vevs
\begin{equation}
\langle \rho \rangle =  H_{\xi\xi}^{(+)}, \quad \langle \rho^\xi \rangle = -2 s_\xi^{(+)},
\end{equation}
\begin{equation}
\langle E^\xi \rangle = H_{tt}^{(+)},
\end{equation}
and
\begin{equation}
\langle O \rangle =  2 s_t^{(+)}.
\end{equation}
Since there is no term in the flux involving $s_t^{(0)}$, we conclude that the vev of $\mathcal E$ vanishes. 
Moreover, this allows us to conclude that the Ward identity for the scaling symmetry is non-anomalous for constant modes. In fact, 
in the absence of any spatial directions and with $z=2$, this Ward identity is just $2 \mathcal E = \mathcal A$, where $\mathcal A$ is the anomaly.
Our flux calculation shows that $\mathcal E = 0$, so it must be the case that  $\mathcal A= 0$. 

The calculation of the flux also enables us to identify possible boundary conditions consistent with flux conservation. An analysis of the inner product \eqref{ip 2} reveals that the leading divergence in \eqref{ip 2} is of order $r^{-5}$ and it involves the modes parametrised by$H_{\xi \xi}^{(-)}$ and $s_{\xi}^{(-)}$. In addition, there are subleading divergences at order $r^{-3}$, $r^{-1}$ and $r^{-1} \log r$ with are proportional to $H_{\xi \xi}^{(+)}$. We conclude then that the modes parametrised by $H_{\xi \xi}^{(-)}$, $H_{\xi \xi}^{(+)}$ and $s_{\xi}^{(-)}$ are non-normalizable. The boundary conditions should fix these modes, and allow the conjugate modes $H_{tt}^{(+)}$, $H_{tt}^{(-)}$ and $s_t^{(+)}$ to vary. Both $H_{t\xi}^{(0)}$ and $s_\xi^{(+)}$ are normalizable, so we can fix either. If we fix $H_{\xi\xi}^{(+)}=0$, we can also fix $s_t^{(0)}$ in a scale invariant way.

\subsubsection{Non-zero $\omega$} 

When we generalise to non-zero $\omega$ for $k_\xi=0$, the structure of the linearised solutions is different from what we might expect. This is because for non-zero $\omega$, the structure of the Ward identities is qualitatively different from the higher-dimensional case. The Ward identities are
\begin{equation}\label{ward w D=3}
\partial_t \mathcal E =0, \quad \partial_t \rho = 0, \quad 2 \mathcal E = \mathcal A,
\end{equation}
where we allow for a non-zero anomaly in the trace Ward identity. Note that because of the absence of spatial directions, $\mathcal E$ and $\rho$ are now just the energy and particle number, rather than densities. The first equation says $\mathcal E$ is a constant. The trace Ward identity will then imply a restriction on the sources: the term appearing in the anomaly for the trace Ward identity must also be a constant.\footnote{Our analysis of the constant modes above found that $\langle \mathcal E \rangle =0$, indicating that the constant part of $\mathcal A$ also vanishes, but this statement is (at least potentially) special to the specific holographic theory we are considering, while the vanishing of the non-constant modes of $\mathcal A$ is a consequence of the general structure of the Ward identities and must be true for any such theory.} Any non-zero $\omega$ modes in $\mathcal A$ must vanish for us to be able to consistently quantise the theory on a given background.  The point is that the Ward identities viewed as a system of linear equations for the non-zero $\omega$ modes of the vevs are degenerate, and so they have no solutions in the inhomogeneous case (with an anomaly source on the right-hand side).

This might seem a remarkably novel feature, but actually the same degeneracy happens for 1+1 dimensional relativistic field theories. There the Ward identities are in general
\begin{equation}
\partial_t T^t_t + \partial_x T^x_t =0, \quad \partial_t T^t_x + \partial_x T^x_x = 0, \quad  T^t_t + T^x_x = \mathcal A \sim R^{(0)},
\end{equation}
where we have noted that the anomaly in this case is proportional to the Ricci scalar of the background geometry. These equations are not generically degenerate, but if we consider the special kinematics where $\omega = \pm k$, that is $\partial_t = \pm \partial_x$, then we have $T^t_t = \mp T^x_t = \pm T^t_x = - T^x_x$, so the left-hand side of the last equation vanishes, and the anomaly contribution must vanish.

This implies that the metric component $h_{uu}$ cannot have a contribution which is just a function of $v$ and independent of $u$, and $h_{vv}$ cannot have a component which is just a function of $u$ and independent of $v$, where $u, v  = t \pm x$ are light-like boundary coordinates. Physically, this is setting some potential non-gauge components of the boundary geometry to zero. In general, the boundary metric in the relativistic 1+1 CFT is pure gauge; by a diffeomorphism and a Weyl transformation one can set the boundary metric to be flat. Working about a background $ds^2 = -2du dv$, the diffeomorphisms and conformal transformation generate a linearised perturbation $h_{uu} = 2 \partial_u \xi_u$, $h_{vv} = 2 \partial_v \xi_v$, $h_{uv} = \partial_{(u} \xi_{v)} + \sigma$. But a component $h_{uu}$ which is independent of $u$ cannot arise from differentiating $\xi_u$, and similarly a component $h_{vv}$ which is independent of $v$ cannot arise from differentiating $\xi_v$, so these modes are not diffeomorphism modes (assuming $x$ is periodically identified, so we do not allow linear functions in $\xi$). But it is precisely these modes that are set to zero by the above anomaly argument, so the theory can only be studied consistently on a background which is in fact diffeomorphic to the flat metric.\footnote{Modulo components which are independent of both $u$ and $v$; these are also not diffeomorphisms, but are not ruled out by the anomaly.}

Returning to the non-relativistic case, this analysis of the Ward identities predicts that for $\omega$ non-zero,  we will have $H_{\xi\xi}^{(+)} = 0$, one restriction on the source modes, and some set of subleading terms involving $\omega^2$. This is precisely what we find. The solution for non-zero $\omega$ is
\begin{align}
\label{Hxixi constant omeganonzero modes D= 3 z=2}
	H_{\xi \xi} &=   0, \\
\label{Htxi constant omeganonzero modes D= 3 z=2}
	H_{t \xi} &=  s_t^{(0)} +  H_{t \xi}^{(0)}  + 2  s_\xi^{(+)} r^2, \\
	H_{tt} &=  - 2 s_t^{(0)}  +
	2 H_{tt}^{(-)}   r^2  - 4  s_\xi^{(+)} (1+2 \log r) r^2  + H_{tt}^{(+)} r^4 - \frac{1}{6} r^6 \omega^2 s_\xi^{(+)}  ,  \\
	s_r &= \frac{i}{2} (s_\xi^{(-)} + s_\xi^{(+)} r^4), \\
	s_t &=  -\frac{1}{3} s_\xi^{(-)} r^{-2} + s_t^{(0)} -  s_\xi^{(+)} r^2 + s_t^{(+)} r^4 +  \frac{1}{4} \omega^2 r^2 s_\xi^{(-)} + \frac{1}{12} \omega^2 r^6 s_\xi^{(+)}, \\
\label{sxi constant omeganonzero modes D= 3 z=2}
	s_\xi &=  s_\xi^{(-)}  r^{-2}    +  s_\xi^{(+)} r^2.
\end{align}
It turns out that the restriction on the sources is to set $H_{\xi\xi}^{(-)} = 0$. This is the source for $\mathcal E^\xi$, which is the extra component in the stress energy complex which is left undetermined because of the degeneration of the Ward identities. It is a non-diffeomorphism mode, as in the above discussion of the relativistic case. 

We can learn more about the structure of the scale Ward identity by looking at the (first order) radial components of the 
equations of motion. More concretely, when we have not yet imposed these first order equations,  the $r^2$ term in 
\eqref{Htxi constant omeganonzero modes D= 3 z=2} appears as an independent constant, which we denote by $H_{t \xi}^{(+)}$. This will of 
course propagate to the other functions, but we do not need the details here. Plugging the solution of the second order equations into the first order equations  we learn that
\begin{align}
\label{ward from asympt D=3 1}
	\omega H_{\xi \xi}^{(+)} &= 0 \\
	\omega ( H_{t \xi}^{(+)} - 2 s_\xi^{(+)} ) &= 0 \\
\label{ward from asympt D=3 2}
	2 (H_{t \xi}^{(+)} - 2 s_\xi^{(+)} ) &=  \frac{1}{2} \omega^2 H_{\xi \xi}^{(-)}
\end{align}
Equations \eqref{ward from asympt D=3 1}-\eqref{ward from asympt D=3 2} correspond to the Ward identities \eqref{ward w D=3}
provided we identify
\begin{equation}
 	\rho \sim H_{\xi \xi}^{(+)} \qquad \mathcal E \sim H_{t \xi}^{(+)} - 2 s_\xi^{(+)} \qquad \mathcal A \sim \omega^2 H_{\xi \xi}^{(-)}
\end{equation} 
\noindent where the $\sim$ indicates equality up to an $\omega$-independent constant. Hence,  
the anomaly is proportional to $\omega^2 H_{\xi \xi}^{(-)}$, and is set to zero due to the conservation equation 
$\omega \mathcal E  = 0$. 

It is also useful to note that the relativistic and Schr\"odinger restrictions are related. If we take the $\bb\to 0$ limit of the Schr\"odinger solution, we recover AdS, and the null coordinate $\xi$ becomes a null coordinate in AdS, so considering zero modes $k_\xi =0$ corresponds in this limit precisely to the special kinematics $\omega = \pm k$ where the AdS Ward identities degenerate, and $H_{\xi\xi}^{(-)} = 0$ reduces to the restriction coming from the Ricci scalar noted above.

\subsection{Non-zero  $k_\xi$}

We now consider the sector of non-zero $k_\xi$. For our purposes, this sector is less interesting, as the bulk modes are just dual to some higher dimension operators in the field theory. However, in previous work on Schr\"odinger as a deformation of AdS, attention has naturally focused on this discussion, as this is the generic kinematics. We will therefore give the full results for purposes of comparison.

For non-zero $k_\xi$, we generally expect the scaling dimensions to depend on $k_\xi$. However, just as in the higher-dimensional case, there are some modes which can be generated by acting with an appropriate $\xi$-dependent diffeomorphism.

For $\omega = 0$, the full solution is
\begin{align}
H_{\xi \xi} &=  H_{\xi \xi}^{(-)} r^{-2} - \frac{1}{2} k_\xi^2 s_t^{(0)}, \\
H_{t \xi} &= - \frac{1}{2} H_{\xi \xi}^{(-)} r^{-2} + H_{t \xi}^{(0)}  - \frac{1}{2} \log r k_\xi^2 s_t^{(0)}, \\
H_{tt} &= - 2 s_t^{(0)} + 4 (1 - \delta_{1} ) s_t^{(1-)} r^{1 - \delta_1} + 4 (1 + \delta_{1} ) s_t^{(1+)} r^{1 + \delta_1},  \\
s_t  &= s_t^{(0)} + s_t^{(1-)} r^{1 - \delta_{1}} +s_t^{(1+)} r^{1 + \delta_{1}}  +s_t^{(3-)} r^{1 - \delta_{3}} +s_t^{(3+)} r^{1 + \delta_{3}}, \\
s_r  &= \frac{i}{2} k_\xi \left( s_t^{(0)} + k_\xi^2 s_t^{(1-)} r^{1 - \delta_1} + k_\xi^2 s_t^{(1+)} r^{1 + \delta_1} -  s_t^{(3-)} r^{1 - \delta_3} -  s_t^{(3+)} r^{1 + \delta_3}  \right), \\
\nonumber
s_\xi &= \frac{1}{4} ( H_{\xi \xi}^{(-)} r^{-2} - k_\xi^2 s_t^{(0)} ) - \frac{1}{2} (3 + \delta_{3}) r^{1-\delta_3} s_t^{(3-)} - \frac{1}{2}(3 - \delta_3)  r^{1+ \delta_3} s_t^{(3+)} \\
			&- \frac{1}{2}(1 - \delta_1) r^{1 - \delta_-} k_\xi^2 s_t^{(1-)} - \frac{1}{2} (1 + \delta_1) k_\xi^2 r^{1 + \delta_1} s_t^{(1+)},
\end{align}
where $\delta_1 = \sqrt{1 + k_\xi^2}$, $\delta_3 = \sqrt{9 + k_\xi^2}$. The bulk diffeomorphism modes are $H_{\xi\xi}^{(-)}$, $H_{t\xi}^{(0)}$ and $s_t^{(0)}$, which correspond to $\delta \hat e^+_\xi$, $\delta \hat e^-_\xi$ and $\delta \hat e^+_t$. As discussed in section \ref{kk}, we can set the first two to zero by gauge-fixing the $\xi$-dependent boundary diffeomorphisms. The logarithmic term does not contribute to the anomaly, as it does not enter at $O(r^2)$, which 
is the right order to modify the vevs and participate in the Ward identity. 

 The other four modes correspond to sources and vevs for scalar operators of dimension
\begin{equation} \label{dd}
 	\Delta = 1 + \sqrt{1 + k_\xi^2}, \qquad \Delta = 1 + \sqrt{9 + k_\xi^2} .
\end{equation}
 These are both irrelevant for all $k_\xi >0$. For general $\omega$, if we set the diffeomorphism modes to zero, we can find the solution for these modes in the same way as in \cite{Guica:2010sw}. We find that we can set $H_{t \xi} = H_{\xi \xi} = 0$
without loss of generality (they only carry diffeo modes), and eliminate $H_{tt}$ and $s_r$ algebraically.
One is left with two coupled second order equations for $s_t$ and $s_\xi$, which can be decoupled by increasing the number of
derivatives. The diff-invariant dynamics are then captured by the following fourth order equation
\begin{align}
\nonumber
	& r^4 s_\xi'''' + 2 r^3 s_\xi''' - (9 + 2 k_\xi^2 + 4 k_\xi \omega r^2) r^2 s_\xi'' + (9 + 2 k_\xi^2 - 4 k_\xi \omega r^2) r s_\xi' \\
	& + [  k_\xi^2 (8 + k_\xi^2) + 4 k_\xi \omega (4 + k_\xi^2) r^2 + 4 k_\xi \omega^2 r^4  ] s_\xi = 0.
\end{align}
The solutions to this equation have the form
\begin{equation}\label{sxi general}
	s_\xi = r^{\Delta} \sum_{i= 0} s_{\xi (i)} r^{2 i},
\end{equation}
where the $s_{\xi (i)}$ are constants and the values of $\Delta$ are those found in the $\omega = 0$ case, corresponding to the source and vev for two operators of dimensions \eqref{dd}.

\subsection{Comparison to previous work}

We now consider the comparison of our results to previous work on null warped AdS$_3$. We focus on the linearised analysis in \cite{Guica:2010sw,vanRees:2012cw} and the analysis of boundary conditions in \cite{Anninos:2010pm}.

Our analysis of the linearised solutions for $k_\xi \neq 0$ is the same as in \cite{Guica:2010sw}. The diffeomorphism modes are what they call the T modes, and the operators of dimension \eqref{dd} correspond to their X modes. The significant difference between our analysis and theirs is our emphasis on the role of the  zero modes. For \cite{Guica:2010sw}, the zero modes are not especially interesting:  T modes are the source for the relativistic stress energy complex, and the zero modes are a special subsector of non-generic kinematics, which they do not consider explicitly. But in our non-relativistic description, the dual field theory lives in one lower dimension, and the zero modes are accordingly the most important sector to understand. We have also seen that the analysis of the zero modes has novel features which do not appear in the discussion for $k_\xi \neq 0$.

The importance of this sector can be stated in a different way that is more independent of our interpretation: Apart from terms determined by the anomalies, the stress tensor only has non-zero components with $k_\xi = 0$. Our perspective focuses in a natural way on this part, and at the price of not being fully covariant in both the $t$ and $\xi$ directions, simplifies the duality by only introducing sources for the potentially non-zero part of the stress tensor at $k_\xi=0$. The $k_\xi \neq 0$ part of the T mode sources considered in \cite{Guica:2010sw} are simply set to zero by gauge-fixing in our approach.

In \cite{vanRees:2012cw}, it was proposed that the appropriate sources for the relativistic stress tensor at $k_\xi \neq 0$ involve a combination of the T and X modes. This does not arise in our analysis. Such a mixing was possible only because the analysis is perturbative in $b$; in our analysis at finite $b$, the diffeomorphism modes and the other modes have different dimensions, so it is not possible for them to mix.

In \cite{Anninos:2010pm}, a  notion of asymptotically Schr\"odinger boundary conditions was proposed, and it was found that the asymptotic symmetry group for these boundary conditions was an infinite extension of the isometry group. Their boundary condition is different from ours, and does not appear to be satisfied bty our linearised solutions. Their analysis was for asymptotically Schr\"odinger rather than asymptotically locally Schr\"odinger boundary conditions, so one might think it should be recovered by  setting the boundary geometry modes in our analysis to zero. However, it is easy to see that our zero mode solutions do not satisfy their boundary conditions in this case. In the constant modes \eqref{Hxixi constant modes D= 3 z=2}, a non-zero $H_{\xi\xi}^{(+)}$ generates a constant metric perturbation $h_{\xi\xi}$. This perturbation violates their boundary conditions, which require that $h_{\xi\xi} \sim \mathcal O(r^2)$ in our notation. In addition, turning on $s_\xi^{(+)}$ will generate a term $h_{tt} \sim s_\xi^{(+)} r^{-2} \log r$, violating their boundary condition $h_{tt} \sim \mathcal O(r^{-2})$.

What if we consider other boundary conditions? In fact $H_{\xi\xi}^{(+)}$ is  a non-normalizable mode, so we should take a boundary condition where it is fixed. If we take $H_{\xi\xi}^{(+)} =0$, this is consistent with the boundary conditions of \cite{Anninos:2010pm}. However, this is fixing the particle number to zero, which seems a strong restriction on the dual field theory. Both $s_\xi^{(+)}$ and its conjugate mode $H_{t\xi}^{(0)}$ are normalizable, so we can choose a boundary condition where $s_\xi^{(+)} = 0$ and $H_{t\xi}^{(0)}$ fluctuates. This has two drawbacks: it's setting the field theory energy to zero, and while we get rid of the problem with $s_\xi^{(+)}$, allowing $H_{t\xi}^{(0)}$ to fluctuate generates $h_{t\xi} \sim H_{t\xi}^{(0)} r^{-2}$, which is again inconsistent with their boundary conditions, which require $h_{t\xi} \sim \mathcal O (r^0)$.

Thus, there is no obvious choice of boundary conditions for our zero modes which will satisfy the assumptions of \cite{Anninos:2010pm}. It would clearly be interesting to analyse the asymptotic symmetries for our boundary conditions defined in section \ref{bc}, but we leave this for future work.

\section{Asymptotic expansion}
\label{asymp}

In this section, we want to go beyond the linearised analysis by showing that solutions of the bulk equations of motion exist for arbitrary boundary data. To do so, we will solve the equations of motion in an asymptotic expansion: that is, we work at large $r$, and solve the equations in an expansion in powers of $r$. Here we restrict ourselves to considering the Fourier zero modes, which include the sources for the stress energy complex, and we will of course be setting sources for irrelevant operators to zero. In the course of demonstrating the existence of this asymptotic expansion, we will also see that when the asymptotic expansion exists we can cancel the divergent terms in the action in the usual way by adding appropriate local counterterms determined by the boundary data.

The general formalism was discussed in \cite{Andrade:2014iia}, but we review it here. We work in terms of the frame fields, and adopt a radial Hamiltonian formalism. The momentum conjugate to $A_\alpha$ is $\pi_\alpha = n^\mu F_{\mu\alpha} = r F_{r\alpha}$. The conjugate to the frame fields is written in terms of a frame extrinsic curvature $K^A_{\ B} = e^\alpha_B \dot e^A_\alpha$, which is not a symmetric object.
 The equations in frame indices are
\begin{align}
\dot K_{(AB)}+ &
K K_{(AB)} + \frac{1}{2}\left(K_{CA}K^C_{\ B}-K_{AC} K_B^{\ C}\right)
+\frac{1}{2}\pi_A\pi_B-\frac{2}{4(d-2)}\eta_{AB} \pi_C\pi^C
\nonumber\\
&=
R_{AB}-\frac{2}{d-2}\Lambda \eta_{AB}-\frac{1}{2}F_{AC}F_B^{\ C} + \frac{2}{8(d-2)}\eta_{AB} F_{CD}F^{CD}-\frac{1}{2}m^2 A_A A_B,
\\
\dot \pi^A+&
K\pi^A-K^A_{\ B}\pi^B = - \nabla_B F^{BA} + m^2 A^A,
\end{align}
and the constraints
\begin{align}
\nabla^A K_{(AB)}-\nabla_B K^A_{\ A} &=
\frac{1}{2} F_{BA} \pi^A + \frac{1}{2}m^2 A_B r A_r,
\\
K^2 -K_{(AB)}K^{AB}-\frac{1}{2}\pi_A \pi^A &=
R-2\Lambda - \frac{1}{4}F_{AB}F^{AB} + \frac{1}{2}m^2 (r A_r)^2 - \frac{1}{2}m^2 A_A A^A, \label{Hconst}
\\
\nabla_A \pi^A &=
 -m^2 r A_r.
\end{align}
Here $F_{AB} = e^\alpha_A e^\beta_B F_{\alpha \beta}$, and $\nabla_A = e^\alpha_A \nabla_\alpha$, where the covariant derivative $\nabla_\alpha$  is a total covariant derivative (covariant with respect to both local Lorentz transformations and diffeomorphisms),  and $\dot {}$ denotes the derivative in the normal direction, which is $-r \partial_r$.


To show that a solution exists in an asymptotic expansion, we want to fix the sources, which will fix the terms appearing on the RHS of these equations, and see that we can satisfy the equations by introducing appropriate subleading terms in $r$ in the expansion which will contribute to the radial derivative terms on the LHS of the equations. For this to work, the source terms need to involve positive powers of $r$. Explicit powers of $r$ enter where there are derivatives along the boundary directions. There are also explicit powers in the Ricci rotation coefficients, determined by $d e^C = \Omega_{AB}^{\ \ \ C} e^A \wedge e^B$.

We restrict ourselves to considering sources which are independent of the $\xi$ coordinate; that is, we assume that the boundary data has a Killing symmetry $\partial_\xi$. Note that we do not assume that $\partial_\xi$ is either null or Killing in the bulk; it is only the boundary sources that are required to have this symmetry, and we can allow the vev modes to be arbitrary functions of $\xi$, this will not affect the derivation of the asymptotic expansion.\footnote{We would not expect it to be possible to extend the construction of an asymptotic expansion to include sources with arbitrary dependence on $\xi$; since the dimensions of the dual operators increase as we increase $k_\xi$, sources with large enough $k_\xi$  are sourcing irrelevant operators, which should cause the expansion to break down. It may be possible to extend the analysis to include sources with sufficiently small $k_\xi$, but as it is not clear what the interesting values might be, we have not attempted to pursue this.} This is thus slightly different from the case considered in \cite{Hartong:2013cba}, where $\partial_\xi$ was taken to be a Killing vector in the bulk.

We need to set to zero the sources for the irrelevant operators. We set the scalar sources $\psi$, $\hat e^+_\xi$ and $\hat e^I_\xi$ to zero by hand. Thus, we assume that
\begin{equation}
\hat e^+ = \hat e^+_a dx^a, \quad \hat e^I = \hat e^I_a dx^a, \quad \hat e^- = \hat e^-_\xi (d\xi + \tilde e^-_a dx^a).
\end{equation}
The one-forms $\hat e^+_a$, $\hat e^I_a$ will then define the boundary geometry the dual field theory lives in, while $\tilde e^-_a$ is a one-form gauge potential (as usual infinitesimal $x^a$ dependent transformations of the $\xi$ coordinate induce gauge transformations of $\tilde e^-_a$), dual to the conserved particle number. Note that because we chose to set the $d\xi$ components of the other frame fields to zero, non-degeneracy implies $\hat e^-_\xi \neq 0$. We will set the source for the irrelevant operator $\mathcal E^i$ to zero without choosing a coordinate system by setting
\begin{equation}
\hat e^+ \wedge d \hat e^+= 0,
\end{equation}
so that the boundary geometry admits a foliation by surfaces of absolute time. For $z=2$, all these restrictions are necessary to ensure the existence of the asymptotic expansion. 

These restrictions on the frame fields imply that the Ricci rotation coefficients
\begin{equation}
\Omega_{+-}^{\ \ \ +} =0, \quad
\Omega_{-I}^{\ \ \ +} =0, \quad
\Omega_{IJ}^{\ \ \ +} =0, \quad \Omega_{-J}^{\ \ \ I} =0, \quad \Omega_{+-}^{\ \ \ I} =0.
\end{equation}
Thus, the non-zero Ricci rotation coefficients are
\begin{equation}
\Omega_{+I}^{\ \ \ +} \sim r, \quad
\Omega_{+J}^{\ \ \ I} \sim r^{2}, \quad
\Omega_{JK}^{\ \ \ I} \sim r,
\end{equation}
\begin{equation}
\Omega_{+-}^{\ \ \ -} \sim r^{2}, \quad
\Omega_{+I}^{\ \ \ -} \sim r^{3}, \quad
\Omega_{-I}^{\ \ \ -} \sim r, \quad
\Omega_{IJ}^{\ \ \ -} \sim r.
\end{equation}
The structure of the one-forms implies $e_-$ has only a $\partial_\xi$ component, so $\partial_-$ vanishes. Thus, the only derivatives appearing are $\partial_+$, which comes with a factor of $r^2$, and $\partial_I$, which comes with a factor of $r$. Thus, we expect an asymptotic expansion to exist for any such boundary data, with arbitrary dependence on $t, x^i$ subject to $\hat e^+ \wedge d \hat e^+= 0$.

This can be checked by analysing the theory in the radial Hamiltonian framework of \cite{Papadimitriou:2004ap,Papadimitriou:2004rz}, as in  \cite{Andrade:2014iia}. This involves expanding in eigenvalues of an appropriate bulk dilatation operator. Assuming that we impose some appropriate boundary or regularity condition in the interior of the spacetime, the on-shell solution of the equations of motion will be uniquely determined in terms of the asymptotic boundary data, so the on-shell action is a function of the boundary data, which we can write as a boundary term,
\begin{equation}\label{Simon50}
S=\int d^{d-1}x \sqrt{-\gamma}\lambda(e^{(A)},\psi).
\end{equation}
We can then think of the canonically conjugate momenta as determined by functional derivatives of this action as in a Hamilton-Jacobi approach, so
\begin{equation}\label{Simon52}
T^{A}_{\ \ B} = \frac{1}{\sqrt{-\gamma}}e_\alpha^{(A)}\frac{\delta}{\delta e_{\alpha}^{(B)}}S,
\end{equation}
\begin{equation}\label{Simon53}
\pi_\psi=\frac{1}{\sqrt{-\gamma}}\frac{\delta}{\delta \psi}S.
\end{equation}
The leading scaling of $\psi$ is $r^{\Delta_-}$, so if we define the dilatation operator
\begin{equation}\label{Simon56}
\delta_D= - \int d^4x \left(
2 e_\alpha^{(+)}\frac{\delta}{\delta e_\alpha^{(+)}}+ e_\alpha^{(I)}\frac{\delta}{\delta e_\alpha^{(I)}}-\Delta_-\psi\frac{\delta}{\delta \psi}
\right),
\end{equation}
then acting on any function of $e^A$, $\psi$, this will agree with the radial derivative at leading order in large $r$, $\delta_D \sim r \partial_r$. Applying this operator to the action, we have
\begin{equation} \label{Simon61}
(d_s+2-\delta_D)\lambda = 2 T^+_{\ph +}+ T^I_{\ph I}-\Delta_- \psi \pi_\psi.
\end{equation}
Compared to the $z<2$ case, we note that \eqref{Simon56} does not involve $e^{(-)}$, and \eqref{Simon61} does not involve $T^-_{\ -}$, as this component does not enter into the trace Ward identity. We determine $\lambda$ by expanding in dilatation eigenvalues and using \eqref{Simon61} and \eqref{Hconst} to determine the contribution at each order in terms of the contributions at earlier orders and the sources.\footnote{As for $z<2$, there is not actually a complete expansion in dilatation eigenvalues, as the logarithms in our linearised solutions indicate that the action of the dilatation operator is not completely diagonalizable. The linearised solution indicates that $\Delta=2$ would be the first order of concern, so these terms contribute at positive powers and thus do not impede the existence of an expansion.} The expansion is
\begin{equation}
\lambda = \sum_{d_s +2 > \Delta\geq 0}\lambda^{(\Delta)} + \ldots, \qquad \delta_D\lambda^{(\Delta)}=\Delta \lambda^{(\Delta)}.
\end{equation}
where $\ldots$ represents terms of higher order which will include logarithms. The dilatation eigenfunctions $\lambda^{(\Delta)}$ are determined by
\begin{align}\label{lambdafromsrcandquad}
(d_s+2-\Delta)\lambda^{(\Delta)} = & -src^{(\Delta)}
\\\nonumber
& +\sum_{s<\Delta/2, s \neq 0} \left[-2K_{(AB)}^{(s)}\pi^{AB(\Delta-s)}-\pi_A^{(s)}\pi^{A(\Delta-s)}-\frac{1}{m^2}(\nabla_A\pi^A)^{(s)}(\nabla_B\pi^B)^{(\Delta-s)}\right]
\\\nonumber
& +\left[-K_{(AB)}^{(\Delta/2)}\pi^{AB(\Delta/2)}-\frac{1}{2}\pi_A^{(\Delta/2)}\pi^{A(\Delta/2)}-\frac{1}{2m^2}(\nabla_A\pi^A)^{(\Delta/2)}(\nabla_B\pi^B)^{(\Delta/2)}
\right].
\end{align}
The quadratic terms in this expression involve lower orders in $\delta$, which are determined from the action by the variations (\ref{Simon52},\ref{Simon53}). We want to focus on the sources:
\begin{equation}
src = R - 2\Lambda -\frac{1}{4}F_{AB}F^{AB}-\frac{m^2}{2}A_A A^A.
\end{equation}
Since we are going to turn $\psi$ off, $A_A A^A=0$, and $F_{AB}$ becomes
\begin{equation}
F_{AB} = 2 \Omega_{AB}^{\ph\ph +} A_+.
\end{equation}
Because of the constraints on the Ricci rotation coefficients, the only non-zero term is $F_{+I}$, so $F^2$ has no non-zero contributions (as $g^{++} =0$). The Ricci scalar is
\begin{equation}\label{Ricciscalar}
R=-4 \partial_A \Omega_{C}^{\ph A C}+\Omega_{CAD}\Omega^{CAD}+2\Omega_{CAD}\Omega^{DAC}+4\Omega_{AD}^{\ph\ph A}\Omega_{C}^{\ph DC}.
\end{equation}
Because of the constraints on the sources, particularly the irrotational condition \eqref{irrotcond}, the Ricci scalar has contributions only at $\Delta =2$. Thus only $-2\Lambda$ contributes to $src^{(0)}$. At $\Delta=2$ we have
\begin{align}
src^{(2)} =
& -4 \partial_+ \Omega_{A-}^{\ph \ph A}-4\partial_- \Omega_{A+}^{\ph \ph A}-4 \partial_I \Omega_{A}^{\ph I A}
+\Omega_{IJK}\Omega^{IJK}
\\\nonumber
& +4\Omega_{+IJ}\Omega_-^{\ph IJ}
+ 4 \Omega_{+I}^{\ph \ph +}\Omega_-^{\ph I-}
\\\nonumber
& + 4 \Omega_{A+}^{\ph \ph B} \Omega_{B-}^{\ph \ph A} + 2 \Omega_{AI}^{\ph\ph B}\Omega_B^{\ph IA}+8\Omega_{A+}^{\ph\ph A}\Omega_{B-}^{\ph\ph B}+4\Omega_{AI}^{\ph\ph A} \Omega_B^{\ph IB},
\end{align}
where $A,B$ are taken to run over $+,-$ and all of the $I$ directions. Since the $\Omega$'s only contribute to sources with positive eigenvalues, we now know a solution for $\lambda$ involving only positive eigenvalues of $\delta_D$ will exist.  As in \cite{Andrade:2014iia}, we could additionally be concerned that $T^+_{\ph -}$, $T^+_{\ph I}$, and $T^I_{\ph -}$ might pick up contributions at negative dilatation eigenvalue from the derivatives of $\lambda$ as in \eqref{Simon52}. However, as discussed in \cite{Andrade:2014iia}, any such contribution would be a boundary scalar, writable entirely in terms of the $\Omega$; as the only nonzero $\Omega$ have positive powers of $r$, the $T^A_{\ph B}$ cannot pick up a contribution at negative eigenvalue. Consequently, the desired asymptotic expansion must exist.

\section{Discussion}
\label{disc}

In this paper, we have constructed the holographic dictionary for Schr\"odinger spacetimes with dynamical exponent $z=2$ based on a frame field formalism. We have proposed a notion of asymptotically locally Schr\"odinger boundary conditions, identified the sources and vevs for the stress tensor complex, and demonstrated that solutions satisfying our boundary conditions exist in an asymptotic expansion. We worked in a theory with a massive vector action in $3$ and $5$ bulk dimensions.  Our method is readily generalizable to other dimensions, and in principle to other supporting matter.

The main difference from our previous analysis of  the Schr\"odinger $z<2$ case in \cite{Andrade:2014iia} is that in the $z=2$ case the $\xi$ direction becomes auxiliary; it is invariant under radial rescalings.  We argued that consequently, as in AdS$_n \times \mathbb R^d$ holography, the appropriate dictionary is formulated by expanding the bulk fields in Fourier modes in $\xi$ and identifying each Fourier mode with the source and vev of a boundary operator $\mathcal O_{k_\xi}$ whose conformal dimension depends on $k_\xi$. This is also different from previous work starting with \cite{Guica:2010sw} which treated $z=2$ Schr\"odinger spacetime as a perturbation of $AdS$. 

In addition, as in the Lifshitz $z=2$ case studied in \cite{Baggio:2011ha,Chemissany:2012du}, there are logarithmic terms in our linearised analysis. In the AdS and Lifshitz cases, the logarithmic terms corresponded to anomalies in the scaling symmetry. However, we find that because of the null structure of the background, some of the logarithms that arise in our case do not contribute to the anomaly at linear order. It would be interesting to understand this from the field theory point of view, or to explore it in the full non-linear theory. We also found a curious feature in the $d_s=0$ case: there is a degeneracy in the Ward identity for the zero modes in $\xi$ that forces us to set source modes that contribute to the scaling anomaly to zero. We noted that a similar feature also appears in the relativistic case for lightlike modes.  

Our analysis followed the philosophy of the work in \cite{Ross:2009ar,Ross:2011gu} on Lifshitz and  \cite{Andrade:2014iia} on Schr\"odinger. We therefore gauge fixed the frame transformation symmetries as much as possible. It would be interesting to explore the analogue of the discussion of Lifshitz in \cite{Christensen:2013lma,Christensen:2013rfa,Hartong:2014pma,Hartong:2014oma} instead, where this symmetry is left unfixed. This has been argued to give a more general perspective on the boundary geometry.

\section*{Acknowledgements}

We are grateful for helpful conversations with Geoffrey Comp\`ere, Monica Guica, Jelle Hartong, Niels Obers, Blaise Rollier, Kostas Skenderis, Marika Taylor, and Balt van Rees. 
The work of AP is supported in part by an STFC studentship. The work of SFR is supported by STFC 
(Consolidated Grant ST/L000407/1). 
The work of TA was supported by the European Research Council under the European Union's Seventh Framework Programme (ERC Grant
agreement 307955).
The work of CK was in part supported by the US Department of Energy under grant DE-FG02-95ER40899, and in part by the Danish Council for Independent Research project ``New horizons in particle and condensed matter physics from black holes''. 
CK and SFR thank the Aspen Center for Physics and NSF Grant \#1066293 for hospitality and support during the completion of this work.
TA thanks the Physics Department of the University of Crete and the Mathematical Sciences Department at University of Southampton 
for their hospitality during the completion of this work.  

\appendix

\section{Spatially dependent modes}
\label{app}

We consider here for completeness the linearised equations for $d_s=2$ with dependence on the spatial directions $\vec x$ included. The equations are the $z \to 2$ limit of the analysis in \cite{Andrade:2014iia}. Considering a single Fourier mode in all boundary directions, we can use the rotation symmetry to orient the spatial coordinates so that the spatial momentum is along the $x$ direction, so the coordinate dependence in all modes is $e^{i \omega t + i k_\xi \xi + i k_x x}$. Then the modes split up into the scalar modes $H_{tt}$, $H_{t\xi}$, $H_{\xi\xi}$, $H_{tx}$, $H_{\xi x}$, $H_{xx}$, $H_{yy}$ $s_t$, $s_x$, $s_r$ and the vector modes $H_{ty}$, $H_{\xi y}$, $H_{xy}$, $s_y$. As in the discussion with no spatial dependence, these all have an expansion in powers of $k_\xi \omega r^2$ and $k_x^2 r^2$. The leading terms take the same form as for the constant modes above.  

The equations of motion in the vector sector are

\begin{align}
\label{E45}
0= &\, rk_x[r^2 \omega H_{\xi y}+k_\xi (H_{ty}+H_{\xi y})]-(k_\xi^2 +2k_\xi\omega r^2)H_{xy}-3r H_{xy}'+r^2 H_{xy}'',
\\
\label{M5}
0= &\, 2 H_{\xi y}-k_x^2 r^2 s_y - (k_\xi^2+2k_\xi\omega r^2 + 5)s_y+r(2H_{\xi y}'-3s_y'+rs_y''),
\\
\label{E35}
0=&\, k_x (k_\xi r H_{xy}-k_x r^2H_{\xi y})+k_\xi^2 H_{ty}-(3+k_\xi\omega r^2)H_{\xi y}+ r(-H_{\xi y}' + rH_{\xi y}''),
\\
\nonumber
0=&\, k_xr(r^{2}\omega H_{xy}-k_x r H_{ty})+(k_\xi\omega r^2 - k_\xi^2 - 5)H_{ty} + 10 s_y + (k_\xi\omega r^2 + \omega^2 r^{4}-2)H_{\xi y}
\\\label{E25}
&\, +r(rH_{ty}''-2(H_{\xi y}+s_y)'-5 H_{ty}'),
\end{align}
and additionally,
\begin{align}
\label{E15}
0=&\, k_\xi [(H_{ty}-H_{\xi y}+2s_y)-r(H_{ty}+H_{\xi y})']-\omega r^{2}[H_{\xi y}+rH_{\xi y}']-k_x r^2 H_{xy}'.
\end{align}

We can solve these equations order by order in $k_\xi \omega$ and $k_x^2$. The subleading components determine the subleading terms in the expansion of the fields. But there are also additional constraints on the leading terms, corresponding to the expected Ward identities. Equation (\ref{E15}) gives at leading order
\begin{equation}
\label{WE15}
k_\xi [ 2H_{ty}^{(+)}-s_y^{(+)}] + 2 \omega H_{\xi y}^{(+)} + 2 k_x \bar{H}_{xy}^{(4)} =0.
\end{equation}
At $k_\xi =0$, this corresponds to the Ward identity
\begin{equation}
\partial_t \mathcal P_y + \partial_x \Pi^x_y =0.
\end{equation}

In the scalar sector, the equations of motion are
\begin{align}
0=&\, 2k_x r(k_\xi  + \omega r^2)(H_{tx}+H_{\xi x})+k_x^2 r^2\Big(\frac{1}{2} H_{tt}-\frac{1}{2}H_{\xi\xi}-H_{t\xi}-k \Big)
+6 H_{tt}-(k_\xi +\omega r^2)^2 k \nonumber
\\\nonumber
&\,  + i (k_\xi  + 2\omega r^2)s_r + 12 s_t  +4 s_\xi - 4rH_{t\xi}' - \frac{1}{2}rH_{\xi\xi}' - 2rk' - rs_\xi ' - r[7H_{tt}' + 2s_t ']+ \frac{1}{2} r^2 H_{tt}''
\\\label{E22}
&\,   + r^2H_{t\xi}'' + \frac{1}{2} r^2 H_{\xi\xi}'' + r^2 k'',
\\
\nonumber
0=&\, k_x r \Big( k_\xi H_{tx} + (2k_\xi +\omega r^2)H_{\xi x} \Big) -k_x^2 r^2(H_{t\xi}+H_{\xi\xi} + k)-2(k_\xi^2 + \omega k_\xi r^2)k
\\\label{E23}
&\,  -3r \Big( H_{t\xi}' + 2k' \Big) + r^2 H_{t\xi}''+ r^2 H_{\xi\xi}'' + 2r^2 k'',
\\
\nonumber
0=&\, rk_x[\omega r^2(H_{\xi\xi} + k)-2i s_r-k_\xi H_{tt}+(\omega r^2-k_\xi )H_{t\xi}] +(k_\xi^2  + k_\xi \omega r^2 - 5)H_{tx}
\\\label{E24}
&\, -(k_\xi\omega r^2+\omega^2 r^{4} - 2)H_{\xi x} -10 s_x + 5 rH_{tx}'+2r(H_{\xi x}' + s_x')-r^2 H_{tx}'',
\\\label{E33}
0=&\,-2 k_x k_\xi r H_{\xi x}+k_x^2 r^2 H_{\xi\xi} + 4 H_{\xi\xi}+2k_\xi^2  k - rH_{\xi\xi}' -r^2 H_{\xi\xi}'',
\\\label{E34}
0=&\,k_x r \Big( k_\xi (H_{t\xi}+k) - \omega r^2 H_{\xi\xi} \Big)-k_\xi^2 H_{tx} +  \Big( k_\xi \omega r^2 +3 \Big) H_{\xi x}+ rH_{\xi x}' - r^2 H_{\xi x}'',
\\
\nonumber
0=&\, r^2 k'' + r^2 H_{\xi\xi}'' + 2 r^2 H_{t\xi}''-rs_\xi '-3rk' -2 r H_{\xi\xi}'-6r H_{t\xi}'+8s_\xi + 2i  k_\xi  s_r - (k_\xi^2  + 2\omega k_\xi r^2)k
\\\label{E44}
&\,  + \Big(-4 + \omega^2 r^{2z} \Big) H_{\xi\xi} - 2k_\xi\omega r^2 H_{t\xi}+k_\xi^2 H_{tt},
\\
\nonumber
0=&\, 2k_x r \Big( k_\xi (H_{tx}+2 H_{\xi x}) + \omega r^2 H_{\xi x} \Big) -r^2 k_x^2 (2H_{tx}+H_{\xi\xi})+r^2 k'' + r^2 H_{\xi\xi}''  -2r s_\xi ' - 3rk'
\\\nonumber
&\,- 2 r H_{\xi\xi}' - 6r H_{t\xi}'+8 s_\xi + 2i k_\xi  s_r  - (k_\xi^2  + 2\omega k_\xi r^2)k + \Big( -4 + \omega^2 r^{4} \Big) H_{\xi\xi} - 2 k_\xi \omega r^2 H_{t\xi}
\\\label{E55}
&\, + k_\xi^2 H_{tt},
\\
\nonumber
0=&\, k_x k_\xi r s_x - k_x^2 r^2 s_\xi + 4H_{\xi\xi} + 2i k_\xi  s_r + k_\xi^2 s_t - \Big( 8+k_\xi \omega r^2 \Big) s_\xi +2r H_{\xi\xi}' - i k_\xi r s_r ' - r s_\xi '
\\\label{M2}
&\,  + r^2 s_\xi '',
\\
\nonumber
0=&\, k_x r(k_\xi  + \omega r^2) s_x - 2 k_x^2 r^2 (s_\xi + s_t)+ 4i (2 k_\xi  + \omega r^2)s_r + 2 \Big( \omega^2 r^{4} - 8 \Big) s_\xi + 2r H_{\xi\xi}'
- 4 r k'
\\\label{M3}
&\,   - 2i (k_\xi  + \omega r^2) r s_r ' - 6r s_\xi ' - 2 k_\xi \omega r^2 s_t - 10 r s_t' + 2r^2 s_t '' + 2r^2 s_\xi '',
\\
\nonumber
0=&\,k_x r \Big( 2i s_r + k_\xi  s_t + (k_\xi  + r^2 \omega)s_\xi - i r s_r ' \Big) + 2 H_{\xi x} - \Big( k_\xi^2  + 2k_\xi \omega r^2 + 5 \Big) s_x
+ 2 r H_{\xi x}'
\\\label{M4}
&  - 3r s_x ' + r^2 s_x '',
\\
\nonumber
\end{align}
and additionally,
\begin{align}
\nonumber
0=&\, k_\xi^2 H_{tt}+2k_\xi k_x r  H_{tx}-2 (k_x^2 + k_\xi \omega)r^2 H_{t\xi}+2rk_x(k_\xi +\omega r^2)H_{\xi x}
-\Big( k_x^2 r^2 + 4 + \omega^2 r^2\Big) H_{\xi\xi}
\\\label{E11}
&\, -\Big(2k_\xi^2 + r^2(k_x^2 + 4k_\xi \omega) \Big) k -2ik_\xi s_r +  8 s_\xi -6rH_{t\xi}'- 4 rH_{\xi\xi}'-6rk'+2 rs_\xi ',
\\
\nonumber
0=&\, k_x r\Big((H_{tx}+H_{\xi x})-rH_{tx}'\Big)+(k_\xi  + 3\omega r^2)H_{\xi\xi}-2k_\xi k -8i s_r-2 \omega r^2 s_\xi
\\\label{E12}
&\,+k_\xi \Big( 2(H_{tt}+s_t)-rH_{tt}' \Big) + (-k_\xi r^{2-z} + \omega r^z)rH_{t\xi}'+\omega r^{z+1}(H_{\xi\xi}'+k'),
\\\label{E13}
0=&\,-k_x r \Big( H_{\xi x}+rH_{\xi x}' \Big)-(k_\xi  + 2\omega r^2)H_{\xi\xi}+k_\xi r H_{t\xi}'-r\omega r^2 H_{\xi\xi}' +2k_\xi r k',
\\
\nonumber
0=&\, k_\xi (H_{tx}-H_{\xi x}+2s_x)-\omega r^2 H_{\xi x}-k_\xi r H_{tx}'-(k_\xi  + \omega r^2)rH_{\xi x}'
\\\label{E14}
&\, +k_x r\Big((H_{\xi\xi}+2s_\xi)+2r(H_{t\xi}'+H_{\xi\xi}'+k' ) \Big),
\\
\nonumber
0=&\,k_x r(4 H_{\xi x}+ 2 s_x + r s_x ') - 2i k_x^2 r^2 s_r + 2(k_\xi  + 2\omega r^2)H_{\xi\xi}- 4 k_\xi k -2i \Big( k_\xi^2 + 8 + 2k_\xi \omega r^2 \Big) s_r
\\\label{M1}
&\,  + 2k_\xi (-2s_t + r s_t ') + 2(k_\xi  + \omega r^2) s_\xi '.
\end{align}
Again, the constraints corresponding to the Ward identities are modified.  Equation (\ref{E14}) gives at leading order
\begin{equation}
\label{WE14}
k_\xi [ 2H_{tx}^{(+)}-s_x^{(+)}] + 2 \omega H_{\xi x}^{(+)} + k_x [ 2k^{(4)} - \frac{5}{3}s_\xi^{(4)}] =0.
\end{equation}
At $k_\xi =0$, this corresponds to the Ward identity
\begin{equation}
\partial_t \mathcal P_x  + \partial_x \Pi^x_x =0. 
\end{equation}
 Equation (\ref{E13}) gives
\begin{equation}
k_\xi \Big( k^{(4)} + \frac{2}{3} s_\xi^{(4)}\Big) - \omega H_{\xi\xi}^{(+)} - k_x H_{\xi x}^{(+)} =0.
\end{equation}
At $k_\xi =0$, this corresponds to the Ward Identity
\begin{equation}
\partial_t \rho + \partial_x \rho^x =0.
\end{equation}
Finally, there is a linear combination of equations (\ref{M1}) and (\ref{E12}) which eliminates $s_r$ giving,
\begin{equation}
k_\xi \Big( 2H_{tt}^{(+)} + s_t^{(+)} \Big) - \omega \Big( 2k^{(4)} - \frac{5}{3} s_\xi^{(4)} \Big)
+ k_x \Big( 2H_{tx}^{(+)} + 2s_x^{(+)} \Big) = 0.
\end{equation}
At $k_\xi =0$, this corresponds to the Ward Identity
\begin{equation}
\partial_t \mathcal E + \partial_x \mathcal E^x =0.
\end{equation}
Thus the full linearised perturbations behave as we expect.

\bibliographystyle{JHEP}
\bibliography{Schrodinger}

\end{document}